\documentclass[preprint,showpacs,preprintnumbers,amsmath,amssymb,nofootinbib]{revtex4}

\usepackage{graphicx}% Include figure files
\usepackage{dcolumn}% Align table columns on decimal point
\usepackage{bm}% bold math

\newcommand{\bi}{\bibitem}
\newcommand{\nn}{\nonumber}

\newcommand{\be}{\begin{eqnarray}}
\newcommand{\ee}{\end{eqnarray}}
\catcode`\@=11
\def\lsim{\mathrel{\mathpalette\@versim<}}
\def\gsim{\mathrel{\mathpalette\@versim>}}
\def\@versim#1#2{\vcenter{\offinterlineskip
\ialign{$\m@th#1\hfil##\hfil$\crcr#2\crcr\sim\crcr } }}
\catcode`\@=12

\begin{document}
\vspace{2cm}
\preprint{KANAZAWA-05-15}

\title{Nonabelian Discrete Family Symmetry \\
to Soften the SUSY Flavor Problem and\\
to Suppress  Proton Decay}

\author{Yuji Kajiyama}
\author{Etsuko Itou}
\author{Jisuke Kubo}

\affiliation{
Institute for Theoretical Physics, Kanazawa
University, Kanazawa 920-1192, Japan
\vspace{3cm}
}

\begin{abstract}
Family symmetry could explain  large mixing
of the atmospheric neutrinos.
The same symmetry could explain why 
the flavor changing current
processes
in  supersymmetric standard models
can be so suppressed. It also may be able to explain
why the proton is so stable.
We investigate these questions in a supersymmetric,
renormalizable
extension of the standard model, which possess a family
symmetry based on a binary dihedral group
$Q_6$. We find that
the amplitude for $\mu \to e+\gamma$
enjoys a suppression factor 
proportional to$ |(V_{\rm MNS})_{e3}|
\simeq m_e/(\sqrt{2}m_\mu)
\simeq 3.4\times 10^{-3}$, and that 
$B(p \to K^0 \mu^+)/B(p\to K^0 e^+)
\simeq 
 |(V_{\rm MNS})_{e3}|^2\simeq 10^{-5}$,
 where $V_{\rm MNS}$ is the neutrino mixing matrix.
 
\vspace{3cm} 
\end{abstract}


\maketitle

\section{Introduction}
The remarkable success of the standard model (SM)  suggests
that we have a highly non-trivial part of a more fundamental
theory for elementary particle physics. 
In spite of this success, the SM suffers from various problems.
One of them is that due to the quadratic divergence 
of the Higgs mass its natural scale can not exceed
$O(1)$ TeV \cite{thooft}.
Therefore,
in order to extend the SM in a natural way, 
the quadratic divergence has to be cancelled \cite{thooft,veltman}.
As it is well known today,
low energy supersymmetry  (SUSY) is introduced to protect
the Higgs mass from the quadratic divergence \cite{susy,ssb}.
Unfortunately,  SUSY is  broken, 
and therefore its  breaking should be soft
to maintain the very nature of low energy SUSY,
whatever its origin is \cite{susy,ssb}.

Renormalizability allows an introduction of a certain set
of soft supersymmetry breaking (SSB) parameters.
In the minimal supersymmetric standard model (MSSM), 
more than 100 SSB  parameters can be introduced \cite{dimopoulos1}.
Moreover,  these parameters should be
highly fine tuned so that they do not
induce unacceptably 
 large flavor changing neutral currents (FCNCs) and
CP violations \cite{fcnc-mueg,fcnc-k,fcnc-edm,fcnc-bsg,fcnc}.
Why should they be so fine tuned?
What controls them ?
This is the so-called SUSY flavor problem, which has existed ever
since SUSY found phenomenological 
applications \cite{dimopoulos2}. 

There are several theoretical approaches
to overcome this problem
\cite{susy,ssb}\cite{gauge}-\cite{maekawa}.
 One of them is to base on a  family symmetry
\cite{hall2}-\cite{maekawa}.
However, if
the family symmetry is hardly 
broken  (i.e. broken by operators with
dimension $\geq 4$), non-symmetric SSB terms
can be generated in higher orders in perturbation theory.
Then it is not possible to
make quantitative statements on 
the induced non-symmetric SSB terms, because these terms
are infinite in perturbation theory, however
small the hard breaking terms are.
In this paper, we consider a supersymmetric model,
recently proposed in  \cite{babu4}, which posses 
a nonabelian discrete family
symmetry based on a binary dihedral group
$Q_{6}$.  We investigate how the SUSY flavor problem
can be softened by $Q_6$ family symmetry,
which is our first task.
We find that except for few cases the fine tuning
of the SSB parameters 
is not necessary in this model.
The only observable process would be
$\mu \to e+\gamma$,
whose branching fraction
enjoys a suppression factor 
proportional to$ |(V_{\rm MNS})_{e3}|^2
\simeq m_e^2/(2m_\mu^2)
\simeq 10^{-5}$, 
and  is three (four)
orders of magnitude larger than that
of $\tau \to e(\mu)+\gamma$ in the model,
where $V_{\rm MNS}$ is the neutrino mixing matrix.

In the MSSM without $R$ parity invariance,
there exist about 100 dimension-four operators 
that violate the baryon number or 
lepton number conservations. 
Fortunately, $R$ invariance can kill all these 
dangerous dimension-four 
operators, while it together with 
$B-L$ conservation allows two types of dimension-five
operators that lead to  proton decay \cite{protondecay, rudaz}.
It is therefore widely expected that these operators are
responsible for an observable proton decay.
Indeed, in the minimal SUSY grand unified theory (GUT)
based on $SU(5)$ \cite{dimopoulos2}, 
these operators force protons to decay faster 
than the experimental bound \cite{pierce}.
Given a family symmetry, the flavor structure is fixed,
at least partially, and it  can control the flavor changing effects 
in  proton decay, too \cite{murayama}.
In this paper we assume that some unknown Planck scale
physics, which respects  $Q_6$ family symmetry,
induces dimension-five operators that lead to  proton decay.
We first  would like to find out
all $B$ and $L$ violating operators
with dimension $\leq 5$ allowed by $Q_6$, and then to
investigate how the family symmetry acts on  proton decay.
If we assume the degeneracy
of the scalar quark masses, then the branching fraction 
$B(p\to K^0~e^+)$ is similar to that of \cite{murayama}
that it becomes comparable with $B(p\to K^+~\bar{\nu})$.
 If the degeneracy assumption is dropped,
$B(p\to K^0~e^+)$ can become 
$O(10^{-4}) \times B(p\to K^+~\bar{\nu})$,
which is still two  orders of magnitude lager than
that of the minimal SUSY GUT case.
It turns out that in contrast to the case of \cite{murayama}
the branching fraction
$B(p\to K^0~\mu^+)$,
which turns out to be  proportional   to $|(V_{\rm MNS})_{e3}|^2$,
 is five orders of magnitude smaller than 
$B(p\to K^0~e^+)$ in the present case, irrespective of the degeneracy
of the scalar quark masses.
The main reason for this  is
the maximal mixing in the charged lepton sector,
which serves the maximal mixing of the
atmospheric neutrinos in this model.
We also give an upper bound on the
coupling constants that are assumed to
be generated by Planck scale  physics.
We find that the family symmetry can raise
the upper bound by two orders of magnitude.

In sect. II we outline the model of \cite{babu4}.
Because of the nonabelian family 
symmetry,
the Yukawa sector of the model is strongly constrained,
and the redundancy of the parameters of this sector is 
significantly reduced. As a consequence,
we can explicitly give
the unitary matrices, $U_L$ and $U_R$,  that rotate the left and
right-handed fermions to diagonalize the corresponding
mass matrices. They are needed to quantitatively discuss
the SUSY flavor problem in the super CKM basis in sect. III.
For completeness, we re-present the 
predictions from $Q_6$ family symmetry.
In sect. IV we consider the $B$ and $L$ violating 
operators that are allowed by the family symmetry
and then calculate the dominant proton decay modes as function
of superpartner masses.
We conclude in sect. V.

\section{The model}

\subsection{Group theory of $Q_{2N}$}
The binary dihedral group $Q_{2N}~
(N=2,3,\dots)$ is a finite subgroup of
$SU(2)$  \footnote{
Models based on 
dihedral flavor symmetries,
ranging from $D3 (\simeq S_3)$ to $Q_6$ and $D_7$,
have been  recently discussed  in
\cite{babu4},  \cite{frampton1}-\cite{ma5}.}.

Its defining matrices are given by
\cite{frampton1,babu4}
\be
\tilde{R}_{2N}&= & 
\left(\begin{array}{ccc}\cos \frac{\theta_N}{2} & \sin 
\frac{\theta_N}{2}
\\ -\sin  \frac{\theta_N}{2} & \cos
 \frac{\theta_N}{2}  \end{array}\right),~
\tilde{P}_Q =
 \left(\begin{array}{ccc}i & 0 \\ 0 & -i
 \end{array}\right),
 \label{PQ}
 \ee
 where $\theta_N=2\pi/N$.
Then the set of $4N$ elements of $Q_{2N}$ is given by
 \be
{\cal G}_{Q_{2N}}=
\{\tilde{R}_{2N}, 
(\tilde{R}_{2N})^2,\dots,(\tilde{R}_{2N})^{2N}={\bf 1},
\tilde{R}_{2N} \tilde{P}_Q ,(\tilde{R}_{2N})^2 \tilde{P}_Q, 
\dots, (\tilde{R}_{2N})^{2N} \tilde{P}_Q=\tilde{P}_Q\}.
\ee
Only one- and two-dimensional irreps. exist.
There exist four different one-dimensional irreps
of $Q_{2N}$, which
can be characterized according to $Z_2\times Z_4$
charge:
\be
{\bf 1}_{+,0}, & &{\bf 1}_{-,0},~{\bf 1}_{+,2},~
{\bf 1}_{-,2} ~~\mbox{for}~~N=2,4,6,\dots,\nn\\
{\bf 1}_{+,0}, & &{\bf 1}_{-,1},~{\bf 1}_{+,2},~
{\bf 1}_{-,3} ~~\mbox{for}~~N=3,5,7,\dots,
\ee
 where the ${\bf 1}_{+,0}$ is the true singlet of $Q_{2N}$,
 and only ${\bf 1}_{-,1}$ and ${\bf 1}_{-,3}$ 
 are complex irreps.
The $N-1$ different two-dimensional irreps are denoted by
\be
{\bf 2}_\ell,  ~~\ell=1,\dots,N-1.
\ee
${\bf 2}_\ell$ with odd $\ell$ is a pseudo real 
representation, while ${\bf 2}_\ell$ with even $\ell$ is a real 
representation.  It is straightforward to calculate the Clebsch-Gordan 
coefficients for tensor products of irreps.
The following  multiplication 
rules in $Q_6$ in particular 
are used to construct the model \cite{babu4}.
\be
%\begin{center}
\begin{array}{ccccccccc}
 {\bf 2}_2  &  \times   
&  {\bf 2}_2  &  =  &  {\bf 1}_{+,0} 
&  +   &  {\bf 1}_{+,2} & + &  {\bf 2}_2 
\\ 
 \left(\begin{array}{c}a_1 \\ a_2  \end{array} \right)   & 
 \times    &  \left(\begin{array}{c}b_1 \\  b_2   \end{array}\right)  
&  =  &   (a_1 b_1 + a_2 b_2)   &  &
 (a_1 b_2 -a_2 b_1)   &    &
 \left(\begin{array}{c}-a_1 b_1 + a_2 b_2 \\ a_1 b_2 +a_2 b_1 \end{array} \right) ,\\ 
\end{array}
%\end{center}
\label{multi3} 
\ee
\be
%\begin{center}
\begin{array}{ccccccccc}
 {\bf 2}_1  &  \times   
&  {\bf 2}_2  &  =  &  {\bf 1}_{-,3}
&  +   &  {\bf 1}_{-,1} & + &  {\bf 2}_1 
\\ 
 \left(\begin{array}{c}x_1 \\ x_2  \end{array} \right)   & 
 \times    &  \left(\begin{array}{c}a_1 \\  a_2 \end{array} \right)  
&  =  &   (x_1 a_2 + x_2 a_1)   &  &
 (x_1 a_1 -x_2 a_2)   &    &
 \left(\begin{array}{c}x_1 a_1 + x_2 a_2 \\ x_1 a_2 -x_2 a_1 \end{array} \right) .  \\ 
\end{array}
%\end{center}
  \label{multi4}
\ee

\subsection{$Q_{6}$ assignment and superpotential }
In Table \ref{assignment} we write the $Q_{6}$ assignment of the quark, 
lepton and Higgs chiral supermultiplets
\footnote{The same model exists for $Q_{2N}$
if $N$ is odd and a multiple  of $3$.}, where
$Q, Q_3, L, L_3$ and  $ H^u, H_3^u, H^d, H_3^d$
stand for $SU(2)_L$ doublets
supermultiplets for quarks, leptons and Higgs bosons, respectively.
Similarly, $SU(2)_L$ singlet
supermultiplets for quarks, charged leptons and neutrinos are denoted by
$U^c, U^c_3,D^c, D^c_3, E^c, E^c_3$ and $N^c, N^c_3$.
$S, T$ and $ Y$ are $SU(3)_C\times SU(2)_L\times U(1)_Y$
singlet Higgs supermultiplets.
\vspace{0.5cm}
\begin{table}
\begin{center}
\begin{tabular}{|c|c|c|c|c|c|c|c|c|c|c|c|c|c|c|}
\hline
 & $Q$ 
 & $Q_3$  
& $U^c,D^c$  
& $U^c_3,D^c_3$ 
 & $L$ & $L_3$ 
 &$E^c,N^c$ & $E_3^c$  &   $N_3^c$ 
  & $H^u,H^d$
 & $H^u_3,H^d_3$ 
 & $S$ & $T$ & $Y$ 
\\ \hline
$Q_6$ &${\bf 2}_1$ & ${\bf 1}_{+,2}$ &
 ${\bf 2}_{2}$ &${\bf 1}_{-,1}$ &${\bf 2}_{2}$ &
${\bf 1}_{+,0}$  & ${\bf 2}_{2}$ & 
${\bf 1}_{+,0}$ & ${\bf 1}_{-,3}$ &
${\bf 2}_{2}$ & ${\bf 1}_{-,1}$  
&${\bf 2}_1$ & ${\bf 2}_{2}$ &
${\bf 1}_{+,2}$\\ \hline
$Z_4$ & $3$ & $3$ &
$0$  & $0$ & 
$1$ & $1$ & $2$ & 
$2$ & $2$  
& $1$ & $1$& $2$ & $2$ &
$0$   \\ \hline
$R$ & $-$ & $-$ &
$-$  & $-$ & 
$-$ & $-$ & $-$ & 
$-$ & $-$  
& $+$ & $+$& $+$ & $+$ &
$+$   \\ \hline
\end{tabular}
\caption{ \footnotesize{$Q_{6}\times Z_4 \times R$ assignment 
of the chiral supermultiplets, where $R$ is the $R$ parity.
This is an alternative assignment to the one given in
 \cite{babu4}. The abelian $Z_4$ is also
 an alterative to the abelian discrete $R$ symmetry  
 $Z_{12R}$ \cite{babu4},
for which one needs more singlets to construct a
desired Higgs sector. 
The anomalies, $Q_6[SU(2)_L]^2,
Q_6[SU(3)_C]^2, Z_4[SU(2)_L]^2$ 
and $ Z_4[SU(3)_C]^2$,
 can be cancelled
by the Green-Schwarz mechanism  if, for instance,
$\kappa_2=\kappa_3$ is satisfied \cite{babu5}, where
$\kappa_2$ and $\kappa_3$ are the Kac-Moody levels 
for $SU(2)_L$ and $SU(3)_C$,
respectively.}}
\label{assignment}
\end{center}
\end{table}
%\normalsize
As an alterative to the abelian discrete $R$ symmetry
$Z_{12R}$   of  \cite{babu4},
for which one needs more singlets to construct a
desired Higgs sector, we introduce an abelian $Z_4$ symmetry.
The $Z_4$ constrains the Higgs sector, where it does not constrain
anything in the Yukawa sector. When discussing  proton decay
in sect. IV , we will  assume  neither $Z_4$ nor $Z_{12R}$,
because the symmetry of the Higgs sector may have a stronger model dependence. 
We then write down the most general, renormalizable,
$Q_{6}\times Z_4 \times R$ invariant superpotential $W$:
\be
W &=& W_Q+W_L+W_H,
\ee
where
\be
W_Q
&=&\sum_{I,i,j,k=1,2,3}\left(
Y_{ij}^{uI} Q_{i} U_{j}^c  H^u_I
+Y_{ij}^{dI} Q_{i} D_{j}^c H^d_I\right),
\label{wQ}\\
W_L
&=&\sum_{I,i,j,k=1,2,3}\left(
Y_{ij}^{eI} L_{i} E_{j}^c  H^d_I
+Y_{ij}^{\nu I} L_{i} N_{j}^c H^u_I\right)
+\frac{m_N}{2} \sum_{i=1,2}(N_i^c)^2
+\frac{\lambda_N}{2} (N_3^c)^2 Y,
\label{wL}\\
W_H &=& m_T (T_1^2+T_2^2)+m_Y Y^2+
\lambda_S (S_1^2+S_2^2)Y\nn\\
& &+\lambda_1 (H_1^u S_2+H_2^u S_1) H_3^d
+\lambda_2 (H_1^d S_2+H_2^d S_1) H_3^u \nn\\
 & &+\lambda_3 \left[ -(H_1^u H_1^d-H_2^u H_2^d) T_1
+ (H_1^u H_2^d+H_2^u H_1^d) T_2\right].
\label{wH}
\ee
The Yukawa matrices $Y$'s are given by
\be
{\bf Y}^{u1(d1)} &=&\left(\begin{array}{ccc}
0 & 0 & 0 \\
0 & 0 & Y_b^{u(d)} \\
0&  Y_{b'}^{u(d)}  & 0 \\
\end{array}\right),~
{\bf Y}^{u2(d2)} =\left(\begin{array}{ccc}
0 & 0 & Y_b^{u(d)}\\
0 & 0 & 0 \\
  -Y_{b'}^{u(d)} &0 & 0 \\
\end{array}\right),\nn\\
{\bf Y}^{u3(d3)}&=&\left(\begin{array}{ccc}
0 & Y_c^{u(d)} & 0\\
Y_c^{u(d)} & 0 & 0 \\
0 &  0 & Y_a^{u(d)} \\
\end{array}\right),
\label{Yuq}
\ee
\be
{\bf Y}^{e1} &=&\left(\begin{array}{ccc}
-Y_c^{e} & 0 & Y_b^{e}\\
0 & Y_c^{e} &  0\\
Y_{b'}^{e}& 0  & 0 \\
\end{array}\right),~
{\bf Y}^{e2} =\left(\begin{array}{ccc}
0 & Y_c^{e} & 0 \\
Y_c^{e} & 0 & Y_b^{e} \\
0&  Y_{b'}^e & 0 \\
\end{array}\right),~
{\bf Y}^{e3}=0,
\label{Yue}
\ee
\be
{\bf Y}^{\nu1} &=&\left(\begin{array}{ccc}
-Y_c^{\nu}& 0 & 0 \\
0 & Y_c^{\nu} & 0 \\
 Y_{b'}^\nu & 0 & 0 \\
\end{array}\right),~
{\bf Y}^{\nu2} =\left(\begin{array}{ccc}
0 & Y_c^{\nu} & 0\\
Y_c^{\nu} &  & 0 \\
0&Y_{b'}^{\nu}  & 0 \\
\end{array}\right),\nn\\
{\bf Y}^{\nu3}&=&\left(\begin{array}{ccc}
0 & 0 & 0\\
0 & 0 & 0 \\
0 &  0 & Y_a^{\nu} \\
\end{array}\right).
\label{Yun}
\ee
All the parameters appearing above are  real, because 
we assume  a spontaneous CP violation to occur.
We will shortly come back to this issue.

The  $Z_4$ charges of the fields are so chosen that the most general 
$Q_6\times Z_{4}\times R$ 
invariant Higgs superpotential (\ref{wH}) 
has an accidental  symmetry:
\be
H_1^{u,d} &\leftrightarrow & H_2^{u,d}, 
S_1\leftrightarrow S_2, T_1\rightarrow -T_1,
\label{parity}
\ee
where $H_3^{u,d}, T_2$ and $Y$ do not transform.
This symmetry 
ensures the stability of the VEV structure
\be
\left\langle H_1^u \right\rangle &=& \left\langle H_2^u
\right\rangle=\frac{1}{2}v^u_D e^{i \theta^u},
~ \left\langle H_1^d \right\rangle = 
\left\langle H_2^d\right\rangle=\frac{1}{2}v^d_D e^{i \theta^d},
~\left\langle H_3^{u,d}\right\rangle
=\frac{1}{\sqrt{2}}v^{u,d}_3 e^{i \theta^{u,d}_3},\nn\\
\left\langle S_1 \right\rangle &=&
\left\langle S_2 \right\rangle
=v^S e^{i (\theta^S/2)},
~\left\langle T_2\right\rangle=v^T e^{i (\theta^T/2)},
~\left\langle Y\right\rangle=v^Y e^{i (\theta^Y/2)},
~\left\langle T_1\right\rangle
=0,~
\label{vev2}
\ee
where $v$'s are non-vanishing real quantities.

The total Higgs potential consists of the supersymmetric part
($D$ terms and $F$ terms) and the $Q_6\times Z_4\times R$
invariant SSB part, where the $B$ and $A$ terms are  given by
\be
{\cal L}_{\rm SSB}^H
&=&
B_T (\tilde{T}_1^2+\tilde{T}_2^2)
+B_Y \tilde{Y}^2+
\lambda_S A_S(\tilde{S}_1^2+\tilde{S}_2^2)\tilde{Y}\nn\\
& &+\lambda_1 A_1(\tilde{H}_1^u \tilde{S}_2+
\tilde{H}_2^u \tilde{S}_1) \tilde{H}_3^d
+\lambda_2 A_2 (\tilde{H}_1^d \tilde{S}_2
+\tilde{H}_2^d \tilde{S}_1) \tilde{H}_3^u \nn\\
 & &+\lambda_3 A_3 
 \left[ -(\tilde{H}_1^u \tilde{H}_1^d
 -\tilde{H}_2^u \tilde{H}_2^d) \tilde{T}_1
+ (\tilde{H}_1^u\tilde{ H}_2^d
+\tilde{H}_2^u \tilde{H}_1^d) \tilde{T}_2\right]+h.c.
\label{LH}
\ee
(The fields with a tilde are the bosonic components
of the corresponding supermultiplets.)
As in the case of the supersymmetric
part, $B$'s and $A$'s are assumed to be real.
We have investigated the minimization conditions for the 
angles $\theta$'s given in (\ref{vev2}), and found that
a nontrivial solution of the conditions can exist.
 How the mass spectrum for this nontrivial solution
looks like is still an open problem. A complete analysis
 of this problem, which is similar to
 that of \cite{okada}, will go beyond
 the scope of the present paper.
We will publish the analysis  elsewhere. 
To proceed, we here simply assume that there
exist a nontrivial CP violating set of VEV's.

\subsection{Fermion mass matrices and 
diagonalization}
 We assume that VEVs take the form (\ref{vev2}), from which
we obtain the fermion mass matrices.
 
 \subsubsection{Quark sector}
 The quark mass matrices are given by
\be
{\bf m}^{u} &=&\frac{1}{2}\left(\begin{array}{ccc}
0 &\sqrt{2} Y_c^{u} v_3^u e^{-i \theta^u_3}
& Y_b^{u}  v_D^u e^{-i \theta^u}\\
\sqrt{2} Y_c^{u}v_3^u e^{-i \theta^u_3} & 0 &
 Y_b^{u} v_D^ue^{-i \theta^u} \\
-Y_{b'}^{u} v_D^ue^{-i \theta^u} & 
Y_{b'}^{u}v_D^u e^{-i \theta^u} & 
\sqrt{2} Y_a^{u}  v_3^u e^{-i \theta^u_3} \\
\end{array}\right),
\label{mu}\\
{\bf m}^{d} &=&\frac{1}{2}\left(\begin{array}{ccc}
0 & \sqrt{2}Y_c^{d} v_3^d e^{-i \theta^d_3}
&Y_b^{d}  v_D^d e^{-i \theta^d}\\
\sqrt{2} Y_c^{d}v_3^d e^{-i \theta^d_3} & 0 &
Y_b^{d} v_D^de^{-i \theta^d} \\
-Y_{b'}^{d} v_D^de^{-i \theta^d} & 
 Y_{b'}^{d}v_D^d e^{-i \theta^d} & 
\sqrt{2}Y_a^{d}  v_3^d e^{-i \theta^d_3} \\
\end{array}\right).
\label{md}
\ee
In the present case, the unitary matrices that rotate
quarks have the following form:
$U_{uL(R)} =  R_{L(R)} P_{L(R)}^u O_{L(R)}^u$,
where $O$'s are orthogonal matrices, and
\be
R_L &=&  \frac{1}{\sqrt{2}}\left( \begin{array}{ccc}
1 & 1 & 0\\-1 & 1& 0\\
 0 & 0 &\sqrt{2}
\end{array}\right),~
R_R =  \frac{1}{\sqrt{2}}\left( \begin{array}{ccc}
-1 & -1 & 0\\-1 & 1& 0\\
 0 & 0 &\sqrt{2}
\end{array}\right),
\label{R}\\
P_L^u &=&    \left( \begin{array}{ccc}
1 & 0 & 0\\0 & \exp (i2\Delta \theta^{u}) & 0\\
 0 & 0&\exp (i\Delta \theta^{u})
\end{array}\right),
\label{Pu}\\
P_R^u &=&
\left( \begin{array}{ccc}
\exp (i2\Delta \theta^{u})  & 0 & 0\\0 & 1& 0\\
 0 & 0&\exp (i\Delta \theta^{u})
\end{array}\right)  
\exp ( i \theta_3^u),
\label{Puc}\\
\Delta \theta^{u} & = &
 \theta_3^u-\theta^u,
 \label{dtheta}
\ee
and similarly for the down sector. The phase matrices $P_{L,R}^{u,d}$
can rotate away the phases of the mass matrices, so that we can bring ${\bf m}^u$ into a real form
\be
{\bf \hat{m}}^u &=& P_L^{u \dag}R^T_L {\bf m}^u R_R P_R^u 
=m_t\left(\begin{array}{ccc}
0 & q_u/y_u & 0  \\ -q_u/y_u & 0 & b_u\\
0 &   b_u'  &  y^2_u \\
\end{array}\right),
\label{muhat}
\ee
which can be then diagonalized as
\footnote{The form of the mass matrix is known as
the next-neighbor interaction form \cite{weinberg,fritzsch1}.}
\be
O^{uT}_L {\bf \hat{m}}^u O_R^u &=&
\left( \begin{array}{ccc}m_u & 0 & \\ 0 & m_c & 0\\ 0 & 0 & m_t
\end{array}\right), 
\ee
and similarly for ${\bf m}^d$.

The CKM matrix $V_{\rm CKM}$ is
 given by 
 \be
 V_{\rm CKM} &=& O^{uT}_L P_L^{u\dag} 
 P_L^{d} O_L^d,
 \label{vckm}
 \ee
 and for the set of the parameters
 \be
\theta_q & = &\theta^d_3-\theta^d-\theta^u_3
+\theta^u  =-1.25,
q_u=0.0002143,
b_u=0.04443,
b'_u=0.09338,\nn\\
y_u &=&0.99732,
q_d =0.005091,
b_d=0.02570,
b'_d=0.77606,
y_d=0.7940,
\label{parameters}
\ee
we obtain
\be
m_u/m_t &=& 1.10\times 10^{-5},
m_c/m_t=4.16 \times 10^{-3},
m_d/m_b=1.22 \times 10^{-3},
m_s/m_b=2.13 \times 10^{-2},
\nn\\
|V_{\rm CKM}| &=&
\left( \begin{array}{ccc}
0.9747 & 0.2236  & 0.0040
\\  0.2234   & 0.9738& 0.0421
  \\ 0.0093 &0.0413& 0.9991
\end{array}\right),~
\sin 2\beta (\phi_1)=0.738.
\label{predQ}
\ee
The experimental values to be compared are \cite{pdg}:
\be
|V_{\rm CKM}^{\rm exp}| &=&
\left( \begin{array}{ccc}
0.9739~\mbox{to}~ 0.9751 & 0.221 
~\mbox{to}~ 0.227 & 0.0029 ~\mbox{to}~ 0.0045
\\  0.221 ~\mbox{to}~ 0.227  & 0.9730
 ~\mbox{to}~ 0.9744   & 0.039 ~\mbox{to}~ 0.044
  \\ 0.0048 ~\mbox{to}~ 0.014  &0.037 
  ~\mbox{to}~ 0.043  & 0.9990 ~\mbox{to}~ 0.9992
\end{array}\right),\nn\\
\sin 2\beta (\phi_1)&=&0.736\pm0.049.
 \ee
 The quark masses at $M_Z$ are given by \cite{kim1}
\be
m_u/m_d &=& 0.541\pm 0.086~(0.61)~,
~m_s/m_d = 18.9\pm 1.6~(17.5),\nn\\
~m_c  &=&0.73\pm 0.17~(0.72)~\mbox{GeV}~,
~m_s =0.058\pm 0.015~(0.062)~\mbox{GeV},\nn\\
~m_t &=& 175 \pm 6 ~\mbox{GeV}~,
~m_b = 2.91\pm 0.07~\mbox{GeV},
\label{qmass}
\ee
where the values in the parentheses are the theoretical values obtained
from  (\ref{predQ}) 
for $m_t= 174  $ GeV and $m_b=2.9  $ GeV. So, we see that the model can well reproduce the experimentally measured 
parameters.
Because of the family symmetry, the CKM parameters and the 
quark masses are related.
In Fig.~1 we plot the predicted area 
 in the $\sin2\phi_1-\phi_3$ plane.
\begin{figure}[htb]
\includegraphics*[width=0.4\textwidth]{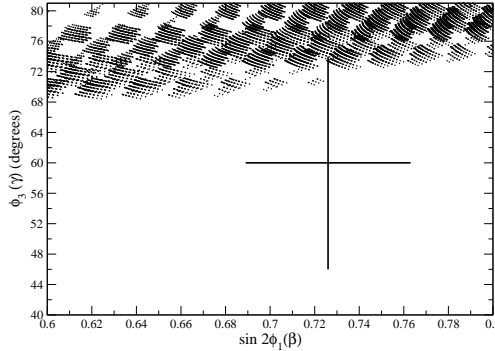}
\caption{\label{fig1}\footnotesize
Predicted area in the $\sin2\phi_1-\phi_3$ plane.
The vertical and horizontal lines correspond to
the experimental values,
$\sin 2 \phi_1(\beta)=0.726\pm 0.037$ and  
$\phi_3=(60^o\pm 14^o) $ \cite{pdg,hfag}.}
\end{figure}
We see from Fig.~1 that the model requires
\be
\phi_3 \gsim 68^o.
\label{pred2}
\ee
Another quantity to be compared may be $|V_{td}/V_{ts}|$,
whose experimental value has been recently obtained from
the observation
of  $b \to d+\gamma$ in the $B$ decays \cite{bdgamma}:
\be
\mbox{Model}&:& 
|V_{td}/V_{ts}| =0.21-0.23,\nn\\
\mbox{Exp.} &:& |V_{td}/V_{ts}|
=0.200\begin{array}{c}+0.026\\-0.025\end{array}
\mbox{(exp.)}
\begin{array}{c}+0.038\\-0.029\end{array}
\mbox{(theo.)}.
\label{pred3}
\ee
We may conclude that 9 independent parameters of the model
can well describe 10  physical observables. 

The mass matrices given in the present model
 can be analytically diagonalized
\cite{harayama1}, and the approximate formula for $V_{\rm CKM}$
 implies that 
   \be
 V_{us} &\simeq &0.794 \sqrt{\frac{m_d}{m_s}}
 -\sqrt{\frac{m_u}{m_c}}e^{-i 2.5},~
  V_{cd} \simeq  -\sqrt{\frac{m_u}{m_c}}
  + 0.794 \sqrt{\frac{m_d}{m_s}}e^{-i 2.5},\nn\\
 V_{cb} &\simeq& 0.81 (\frac{m_s}{m_b})e^{ -i 2.5}
 - 9.63(\frac{m_c}{m_t})e^{ -i 1.25},
  \label{vcb}
 \ee
which should be compared with the Fritzsch formulas
 \cite{fritzsch1}.

Finally, we give the unitary matrices that rotate the quarks
for the choice of the parameters given in (\ref{parameters}):
\be
U_{uL(R)} &=&R_{L(R)}^u P_{L(R)}^uO_{L(R)}^u,
~U_{dL(R)} =R_{L(R)}^d P_{L(R)}^dO_{L(R)}^d,
\label{Uq}
\ee
where $R_{L,R}^u$ and $R_{L,R}^d$ are given in 
(\ref{R}) and  (\ref{Pu}), and
\be
O_L^u &\simeq &
\left(\begin{array}{ccc}
0.9987 & 0.0514 &2\times 10^{-5}\\
0.0513 & -0.9977 & 0.0442\\
-0.0023 & 0.0442 & 0.9990
\end{array}\right),~
O_R^u \simeq 
\left(\begin{array}{ccc}
-0.9987 & 0.0515 & -10^{-5}\\
0.0513 & 0.9943& 0.0933\\
 -0.0048& -0.0932& 0.9956
\end{array}\right),
\label{Ou}\\
  O_L^d &\simeq &
\left(\begin{array}{ccc}
0.9834 & 0.1814 & 0.0050\\
0.1813&  -0.9833& 0.0162\\
-0.0078&  0.0150& 0.9999
\end{array}\right),~
  O_R^d \simeq 
\left(\begin{array}{ccc}
-0.9552 & 0.2960 & -0.0001\\
0.1866 &  0.6025 &  0.7760\\
 -0.2298 & -0.7412 & 0.6308
  \end{array}\right).
  \label{Od}
\ee
Using these orthogonal matrices and 
the matrices defined in (\ref{R}) $\sim$ (\ref{dtheta}), 
Eq. (\ref{Uq}) gives 
the unitary matrices in the explicit form:
\be
U_{uL}&=&\left(\begin{array}{ccc}
0.706  & 0.0363 &  1.42 \times 10^{-5} \\
-0.706  & -0.0363  & -1.42 \times 10^{-5} \\
0 & 0 &  0\\
\end{array}\right), \nn \\ 
&+&e^{2 i \Delta \theta^u}\left(\begin{array}{ccc}
0.0363 & -0.705 & 0.0313 \\
0.0363 & -0.705 & 0.0313 \\
-0.00229 ~e^{-i \Delta \theta^u} & 0.0442~ e^{-i \Delta \theta^u} & 
0.999~ e^{-i \Delta \theta^u} \\
\end{array}\right), \label{UuL} \\ 
U_{dL}&=&\left(\begin{array}{ccc}
0.695  & 0.128 &  0.00352 \\
-0.695  & -0.128  & -0.00352 \\
0 & 0 &  0\\
\end{array}\right), \nn \\ 
&+&e^{2 i \Delta \theta^d}\left(\begin{array}{ccc}
0.128 & -0.695 & 0.0115 \\
0.128 & -0.695 & 0.0115 \\
-0.00783 ~e^{-i \Delta \theta^d} & 0.0150~ e^{-i \Delta \theta^d} & 
1.00~ e^{-i \Delta \theta^d} \\
\end{array}\right), \label{UdL} \\  
U_{uR}&=&e^{i \theta^u_3}\left(\begin{array}{ccc}
0.0363  & 0.703 &  0.0660 \\
0.0363  & 0.703  & 0.0660 \\
0 & 0 &  0\\
\end{array}\right)  \nn \\ 
&+&e^{2 i \Delta \theta^u+i \theta^u_3}\left(\begin{array}{ccc}
-0.706 & 0.0364 & -6.72 \times 10^{-6} \\
0.706 & -0.0364 & 6.72  \times 10^{-6} \\
-0.00482 ~e^{-i \Delta \theta^u} & -0.0932~ e^{-i \Delta \theta^u} & 
0.996~ e^{-i \Delta \theta^u} \\
\end{array}\right), \label{UuR} \\ 
U_{dR}&=&e^{i \theta^d_3}\left(\begin{array}{ccc}
0.132  & 0.426 &  0.549 \\
0.132  & 0.426  & 0.549 \\
0 & 0 &  0\\
\end{array}\right)  \nn \\ 
&+&e^{2 i \Delta \theta^d+i \theta^d_3}\left(\begin{array}{ccc}
-0.675 & 0.209 & -7.35 \times 10^{-5} \\
0.675 &- 0.209 & 7.35 \times 10^{-5} \\
-0.230 ~e^{-i \Delta \theta^d} &- 0.741~ e^{-i \Delta \theta^d} & 
0.631~ e^{-i \Delta \theta^d} \\
\end{array}\right). \label{UdR}
\ee
The unitary matrices above will be used 
when discussing the SUSY flavor problem in sect. III
and  proton decay in sect. IV.

\subsubsection{Lepton sector}
The mass matrices  in the lepton sectors are:
\be
{\bf m}^{e} &=&\frac{1}{2} \left(\begin{array}{ccc}
- Y_{c}^{e}& 
Y_{c}^{e}
&  Y_b^{e} \\
 Y_{c}^{e} & 
  Y_{c}^{e} &
 Y_b^{e}  \\
Y_{b'}^{e}  & 
Y_{b'}^{e} & 
0 \\
\end{array}\right)  v_D^d e^{-i \theta^d},
\label{me}\\
{\bf m}^{\nu} &=&\frac{1}{2} \left(\begin{array}{ccc}
- Y_{c}^{\nu}& 
Y_{c}^{\nu}
&0 \\
Y_{c}^{\nu} & 
Y_{c}^{\nu} &0  \\
Y_{b'}^{\nu}  & 
Y_{b'}^{\nu} & 
\sqrt{2}  Y_{a}^{\nu}\tan\gamma^u e^{i (\theta^u_3-\theta^u)} \\
\end{array}\right)v_D^u  e^{i \theta^u},
\label{mn}
\ee
where $\tan\gamma^u=v_3^u/v_D^u$.
We start with the mass matrix of the charged leptons ${\bf m}^e$:
\be
U_{eL}^{\dag}{\bf m}^{e} U_{eR} &=&
\left( \begin{array}{ccc}m_e & 0 & \\ 0 &
 m_\mu & 0\\ 0 & 0 & m_\tau
\end{array}\right).
\ee
One finds \cite{kobayashi3} that  $U_{eL}$ and $ U_{eR}$ can be approximately 
written
as
\be
U_{eL} &= &R\left(
\begin{array}{ccc}
-\epsilon_e( 1 + \epsilon_\mu^2) &
-(1/\sqrt{2})(1-\epsilon_e^2-\epsilon_e^2 \epsilon_\mu^2 ) &
1/\sqrt{2} \\
\epsilon_e( 1 - \epsilon_\mu^2) &
(1/\sqrt{2}) (1-\epsilon_e^2+\epsilon_e^2 \epsilon_\mu^2) &
1/\sqrt{2}\\
1-\epsilon_e^2
&- \sqrt{2}\epsilon_e &  \sqrt{2} \epsilon_e \epsilon_\mu^2
\end{array}\right),
\label{UeL}
\\
U_{eR} &= &R\left(
\begin{array}{ccc}
1 -\epsilon_\mu^2/2 & -\epsilon_e^2(1-\epsilon_\mu^2)
 & \epsilon_\mu \\
 -\epsilon_e^2(1-\epsilon_\mu^2/2) & -1  & 0 \\
-\epsilon_\mu & \epsilon_e^2 \epsilon_\mu &  1 - \epsilon_\mu^2/2
\end{array}\right)e^{i \theta^d},
\label{UeR}
\ee
where $R$  interchanges the first and second row, 
\be
R=\left( \begin{array}{ccc}
0 & 1 & 0 \\
1 & 0 & 0 \\
0 & 0 & 1 \\
\end{array}\right), 
\label{Rlep}
\ee
and $\epsilon_\mu =m_{\mu}/ m_{\tau}$ and
$\epsilon_e =m_e/(\sqrt{2}m_{\mu})$.
In the limit $m_e=0$, the unitary matrix $U_{eL}$
becomes
$$
\left(
\begin{array}{ccc}
0&1/\sqrt{2} &1/\sqrt{2}\\
0 &-1/\sqrt{2}  &1/\sqrt{2}\\1& 0 &  0
\end{array}\right),
$$
which is the origin for a maximal mixing of
the atmospheric neutrinos.

As for the neutrino masses, we assume that
a see-saw mechanism \cite{seesaw} takes place.
As we can see from (\ref{wL}), 
the mass matrix for the right-handed
neutrinos is diagonal:
${\bf m}_N=(
(-1)^\eta m_N, (-1)^\eta m_N, \lambda_N v^Y 
\exp (i\theta_Y/2)),~(\eta=0~\mbox{or}~1)$,
where $m_N$ and $ \lambda_N  v^Y$ are real and positive  by assumption.
Therefore, the Majorana mass matrix
for the left-handed neutrinos becomes
\be
{\bf M}_\nu
&=&{\bf m}^{\nu} {\bf m}_N^{-1}({\bf m}^\nu)^T\nn\\
&=& (-1)^\eta e^{i 2\theta^u}R \left( \begin{array}{ccc}
2 (\rho_{2})^2 & 0 & 
2 \rho_2 \rho_{4}
\\ 0 & 2 (\rho_{2})^2 & 0
  \\ 2 \rho_2 \rho_{4} & 0  &  
2 (\rho_{4})^2 +
 (\rho_3)^2\exp i 2 \varphi_{3}
\end{array}\right) R,
\label{m-nu}
\ee
where $R$ is given in (\ref{Rlep}), and 
\be
2\varphi_3 &=&2(\theta_3^u-\theta^u)+\eta \pi
-\theta^Y/2,
\label{phi3}\\
\rho_{2} &=&
\frac{1}{2}Y_c^{\nu} v_D^u /\sqrt{m_N}~,~
\rho_4 = 
\frac{1}{2}Y_{b'}^{\nu} v_D^u  /\sqrt{m_N},~
\rho_3 =\frac{1}{\sqrt{2}}Y_a^{\nu} v_3^u  /\sqrt{ Y_N v^Y}.
\label{rhos}
\ee
The  factor $(-1)^\eta e^{i 2\theta^u}$ has
no effect, and so we ignore it in the following discussions.

Noticing that the $\rho$'s in (\ref{m-nu}) are real numbers,
we recall that that ${\bf M}_{\nu}$ can be diagonalized as \cite{kubo1}
\be
U^T_\nu {\bf M}_{\nu} U_\nu &=& \left( \begin{array}{ccc}
m_{\nu_1} & 0 & 0\\
0 & m_{\nu_2} &0 \\
0 & 0 & m_{\nu_3}
\end{array}\right),
\ee
where $c_{12}=\cos\theta_{12}$, and $s_{12}=\sin\theta_{12}$, and
\be
U_{\nu}&= &R\left( \begin{array}{ccc}
-s_{12}e^{i(\phi_\nu-i\varphi_1)/2} & c_{12}e^{i(\phi_\nu-i\varphi_2)/2}
&  0
\\ 0 & 0 &-1
\\    c_{12}e^{-i(\phi_\nu+i\varphi_1)/2}  
& s_{12} e^{-i(\phi_\nu+i\varphi_2)/2}& 0
 \end{array}\right),
\label{unumax3}\\
m_{\nu_3} \sin \phi_\nu &=& m_{\nu_2} \sin \varphi_2
=m_{\nu_1} \sin \varphi_1~,~2 \varphi_{3}=\varphi_{1}+\varphi_{2},
\label{sinp}\\
\frac{m_{\nu_2}^2}{\Delta m_{23}^2}
&\simeq &
\frac{1}{\sin^2 2\theta_{12}\cos^2 \phi_\nu}
-\tan^2 \phi_\nu ~~\mbox{for}~~|r| << 1.
\label{mnu21}
\ee
It turns out that  only an inverted mass spectrum
\be
m_{\nu_3} & < & m_{\nu_1}, m_{\nu_2}
\label{spectrum}
\ee
is consistent with  the experimental constraint $ |\Delta m_{21}^2|
< |\Delta m_{23}^2|$  in the present model.
Note that Eq. (\ref{sinp}) is satisfied for
\be
2 \varphi_{3} &=& \varphi_{1}+\varphi_{2} \sim \pm\pi
\label{1plus2}
\ee
and not for $\varphi_{1} \sim \varphi_{2}$. 

The mixing matrix  $V_{\rm MNS}$
can be   obtained from $U_{eL}^{\dag} U_\nu$,
where $U_{eL}$ and $ U_\nu$ are given in 
(\ref{UeL}) and (\ref{unumax3}), respectively.
 The product $U_{eL}^{\dag} U_\nu$
 can be brought by an
appropriate  phase transformation  to
a popular form, which in the present model approximately assumes the form
\be
V_{\rm MNS}  &\simeq & 
\left( \begin{array}{ccc}
 c_{12}  & s_{12}  &   s_{13} e^{-i\delta}\\
  -s_{12} /\sqrt{2}   & 
   c_{12} /\sqrt{2}   &   1 /\sqrt{2} \\
 s_{12} /\sqrt{2} & 
   -c_{12} /\sqrt{2}    &   1 /\sqrt{2}
\end{array}\right) 
 \times
\left( \begin{array}{ccc}
1 & 0 & 0\\
0 & e^{i \alpha} &0 \\
0 & 0 & e^{i \beta}
\end{array}\right).
\label{vmns}
\ee
with
\be
s_{13} & \simeq &\epsilon_e=\frac{m_e}{\sqrt{2}m_\mu}
=3.4\cdots\times 10^{-3},~\delta_{CP}\simeq -\phi_\nu,
\label{sin13}\\
\sin 2 \alpha &=&\sin(\varphi_1-\varphi_2),~
\sin 2 \beta =\sin(\varphi_1-\varphi_\nu),
\label{sinb}
\ee
where $\varphi_1,\varphi_2$ and $\phi_\nu$ are defined in (\ref{sinp})
\footnote{Unfortunately, this value of $s_{13}$ is too small
to be measured \cite{minakata}.}. 
There are seven independent parameters to describe 
12 parameters ($3+3=6$ masses, three angles and three phases)
of the lepton sector.
Therefore, the effective Majorana mass $<m_{ee}>$ in neutrinoless
double $\beta$ decay, for instance,  can be predicted.
In Fig.~2 we  plot $<m_{ee}>$  as a function of $\sin \phi_\nu$
for $\sin^2\theta_{12}=0.3, \Delta m_{21}^2=6.9 \times10^{-5}$ eV$^2$ and
$\Delta m_{23}^2=1.4, 2.3, 3.0 \times 10^{-3}$ eV$^2$ \cite{maltoni3}.
As we  see from Fig.~2, 
the prediction  is consistent with recent experiments \cite{klapdor1,wmap}.

We will use the unitary matrices given in
(\ref{UeL}), (\ref{UeR}) and  (\ref{unumax3})
in sect.   III and IV, and  will see that the small parameter
$s_{13}=\epsilon_e=m_e/(\sqrt{2})/m_\mu
\simeq 0.0034$ appears as a suppression factor
in FCNCs as well as in some of  proton decay modes.

\begin{center}
\begin{figure}[htb]
\includegraphics*[width=0.6\textwidth]{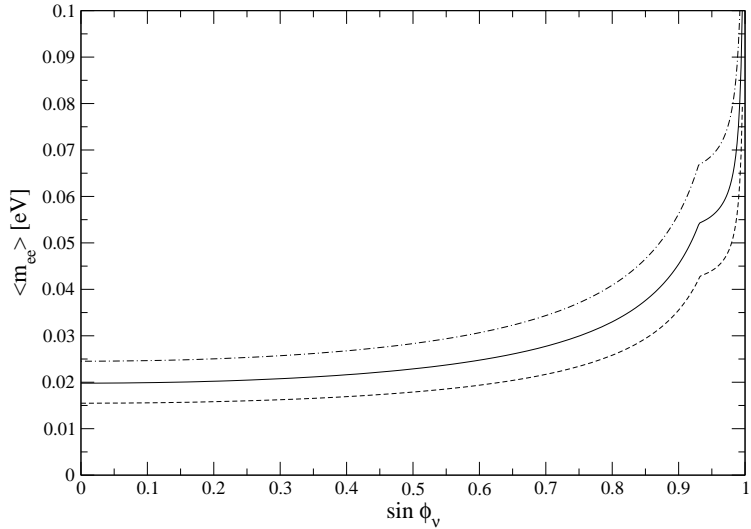}
\caption{\label{fig2}
The effective Majorana mass $<m_{ee}>$ as a function of
$\sin \phi_\nu$ with 
$\sin^2\theta_{12}=0.3$ and 
 $\Delta m_{21}^2=6.9 \times10^{-5}$ eV$^2$.
 The dashed, solid and dot-dashed lines stand for 
$\Delta m_{23}^2=1.4, 2.3$ and $ 3.0 \times 10^{-3}$ eV$^2$,
respectively. The $\Delta m_{21}^2$ dependence is very small.}
\end{figure}
\end{center}

\section{The SUSY flavor problem}
If three generations of a family is put into
a one-dimensional and two-dimensional irreps of 
any dihedral group, then 
the soft scalar mass matrix for the sfermions has always a diagonal form: 
\be
{\bf \tilde{m}^2}_{(q,\ell)LL} =
{m}^2_{\tilde q,\tilde \ell} \left(
\begin{array}{ccc}
a_L^{q,\ell} & 0 & 0 \\
0 & a_L^{q,\ell} & 0 \\
0 & 0 & b_L^{q,\ell}
\end{array}
\right),~
{\bf \tilde{m}^2}_{aRR} =
{m}^2_{\tilde q,\tilde \ell} \left(
\begin{array}{ccc}
a_R^{a} & 0 & 0 \\
0 & a_R^{a} & 0 \\
0 & 0 & b_R^{a}
\end{array}
\right)~~~(a=u,d,e),
\label{scalarmass}
\ee
where ${m}_{\tilde q,\tilde \ell}$ denote the average of the  squark 
and slepton masses, respectively,  and $(a_{L(R)}, b_{L(R)})$ are
dimensionless free real parameters of $O(1)$.
Further, since the trilinear  interactions ($A$ terms)
are also  $Q_6$ invariant,
the left-right mass matrices assume the form
\be
\left({\bf \tilde{m}^2}_{aLR}\right)_{ij} 
&=&
A_{ij}^a\left( {\bf m}^a \right)_{ij} 
~~(a=u,d,e),
\label{Aterm}
\ee
where $A_i^{a}$'s are free parameters of dimension one,
and the fermion masses ${\bf m}$'s are given in
(\ref{mu}), (\ref{md}) and (\ref{me}).
Here we 
assume that $A_i^{a}$'s are in  the same order as the gaugino masses.
They are real, because we impose CP invariance at
the Lagrangian level.

We work in the super CKM basis and 
calculate
\be
\Delta_{LL}^{a} &=&
U_{aL}^{\dagger} ~{\bf \tilde{m}^2}_{(q,\ell)LL}~
 U_{aL}~\mbox{and}~
\Delta_{RR(LR)}^{a} =
U_{aR(L)}^{\dagger}~ {\bf \tilde{m}^2}_{aRR(LR)} ~U_{aR}
\label{Delta1}
\ee
to parameterize FCNCs and CP violations coming
from the SSB sector, where the unitary matrices 
$ U$'s are given in (\ref{Uq}), (\ref{UeL})
and (\ref{UeR}).
In doing so, one observes that something interesting happens;
a phase alignment.
This is because the only source for  CP phases comes from VEVs
(\ref{vev2}). To see the
phase alignment, we first observe that
the matrices $R_{L,R}$
and the phase rotation matrices
$P_{L,R}^{u,d}$, given in (\ref{R}), (\ref{Pu}) and (\ref{Puc}),  
commute with 
the scalar soft mass matrices 
${\bf \tilde{m}}^2_{qLL}$ and ${\bf \tilde{m}}^2_{u,dRR}$.
This implies that $\Delta_{LL,RR}^{u,d} $ are real,
where $\Delta_{LL,RR}^{e} $ is trivially real
as we see from (\ref{UeL}) and (\ref{UeR}).
As for the left-right soft mass squared (\ref{Aterm}),
we find that
$(P_L^a)^{\dagger}~(R^a_L)^T
 {\bf \tilde{m}^2}_{aLR} ~R^a_R P_R^a
$ is a real matrix for all $a=u,d,e$.
Consequently, no $CP$ violating processes
induced by the SSB terms are possible in this model,
satisfying the most stringent experimental constraint
coming from the EDM of the neutron and the electron \cite{fcnc}.

In \cite{fcnc}, experimental bounds  on the dimensionless quantities
\be
\delta^{a}_{LL,RR,LR} &=& 
\Delta^{a}_{LL,RR,LR}/{m}^2_{\tilde q,\tilde \ell}~~~(a=u,d,e),
\ee
are given.
The theoretical values of $\delta$'s for the present model 
are calculated below,
where 
\be
\Delta a_{L}^{q,\ell} &=&a_{L}^{q,\ell}-b_{L}^{q,\ell},~
\Delta a_{R}^{a} =a_{R}^{a}-b_{R}^{a},~
\tilde{A}_{i}^{a} = \frac{A_{i}^{a}}{{m}_{\tilde q,\tilde \ell}}~~(a=u,d,e),
\label{deltaAt1}
 \ee
are introduced.

\vskip 0.5cm
\underline{\bf Leptonic sector ($LL$ and $RR$): }
%\vskip 0.5cm
\be
(\delta^{e}_{12})_{LL}
&=& (\delta^{e}_{21})_{LL}\simeq
 4.9 \times 10^{-3} ~\Delta a_L^{\ell}, \nn\\
(\delta^{e}_{13})_{LL}
&=&  (\delta^{e}_{31})_{LL}
\simeq -1.7 \times 10^{-5} ~\Delta a_L^{\ell}, \nn\\
(\delta^{e}_{23})_{LL}
&=& (\delta^{e}_{32})_{LL}
\simeq  8.4 \times 10^{-8} ~\Delta a_L^{\ell},\nn\\
(\delta^{e}_{12})_{RR}
&=& (\delta^{e}_{21})_{RR}
\simeq 8.4 \times 10^{-8} ~\Delta a_R^{e}, \label{deltaell}
\\
(\delta^{e}_{13})_{RR}
&=& (\delta^{e}_{31})_{RR}
\simeq 5.9 \times 10^{-2} ~\Delta a_R^{e}, \nn\\
(\delta^{e}_{23})_{RR}
&=& (\delta^{e}_{32})_{RR} 
\simeq -1.4 \times 10^{-6} ~\Delta a_R^{e}.\nn
\ee

\vskip 0.5cm
\underline{\bf Leptonic sector ($LR$): }
%\vskip 0.5cm
\be
(\delta^{e}_{12})_{LR}
&\simeq& 5.1 \times 10^{-6} ~
\left(\tilde{A}_c^{e}-\tilde{A}_{b'}^{e}\right)
\left(\frac{100~\mbox{GeV}}{m_{\tilde{\ell}} }\right), \nn\\
(\delta^{e}_{21})_{LR}
&\simeq& 2.5 \times 10^{-8} ~
\left(\tilde{A}_c^{e}-\tilde{A}_{b'}^{e}\right)
\left(\frac{100~\mbox{GeV}}{m_{\tilde{\ell}} }\right), 
\label{LRell}\\
(\delta^{e}_{13})_{LR}
&\simeq& 3.1 \times 10^{-7} ~
\left(\tilde{A}_{b'}^{e}-\tilde{A}_b^{e}\right)
\left( \frac{100~\mbox{GeV}}{m_{\tilde{\ell}} }\right), \nn\\
(\delta^{e}_{31})_{LR}
&\simeq& 1.1 \times 10^{-3} ~
\left(\tilde{A}_c^{e}-\tilde{A}_b^{e}\right)
\left( \frac{100~\mbox{GeV}}{m_{\tilde{\ell}} }\right), \nn\\
(\delta^{e}_{23})_{LR}
&\simeq&  -1.5 \times 10^{-9} ~
\left(\tilde{A}_{b'}^{e}-\tilde{A}_b^{e}\right)
\left( \frac{100~\mbox{GeV}}{m_{\tilde{\ell}} }\right),\nn\\
(\delta^{e}_{32})_{LR}
&\simeq& -2.5 \times 10^{-8} ~
\left(\tilde{A}_c^{e}-\tilde{A}_b^{e}\right)
\left( \frac{100~\mbox{GeV}}{m_{\tilde{\ell}} }\right).\nn
\ee

\vskip 0.5cm
\underline{\bf Up quark  sector ($LL$ and $RR$):}
%\vskip 0.5cm
\be
(\delta^u_{12})_{LL}
&=&  (\delta^u_{21})_{LL}\simeq 
1.0 \times 10^{-4}~\Delta a_L^{q}, \nn\\
(\delta^u_{13})_{LL}
&=& (\delta^u_{31})_{LL}
\simeq  2.3\times 10^{-3}~\Delta a_L^q, \nn\\
(\delta^u_{23})_{LL}
&=& (\delta^u_{32})_{LL}
\simeq  -4.4 \times 10^{-2}~\Delta a_L^{q},\nn\\
(\delta^u_{12})_{RR}
&=& (\delta^u_{21})_{RR}
\simeq -4.5 \times 10^{-4} ~\Delta a_R^{u}, \label{deltau}
\\
(\delta^u_{13})_{RR}
&=& (\delta^u_{31})_{RR}
\simeq 4.8 \times 10^{-3}~\Delta a_R^{u}, \nn\\
(\delta^u_{23})_{RR}
&=& (\delta^u_{32})_{RR}
\simeq 9.3 \times 10^{-2}~\Delta a_R^{u}.\nn
\ee

\vskip 0.5cm
\underline{\bf Up quark  sector  ($LR$)}:
%\vskip 0.5cm
\be
(\delta^{u}_{11})_{LR}
&\simeq& 10^{-6}
\left[3.8 \left( \tilde{A}_{a}^{u}-\tilde{A}_{b}^{u}-\tilde{A}_{b'}^{u} +2 \tilde{A}_c^u\right)~\right] 
\left( \frac{500 ~\mbox{GeV}}{m_{\tilde{q}} }\right), \nn\\
(\delta^{u}_{22})_{LR}
&\simeq& 10^{-3}
\left[ -1.4  \left( \tilde{A}_{a}^{u}-\tilde{A}_{b}^{u}-\tilde{A}_{b'}^{u} \right)
+7.7 \times 10^{-3} \tilde{A}_{c}^{u}~\right] 
\left( \frac{500 ~\mbox{GeV}}{m_{\tilde{q}} }\right), \nn\\
(\delta^{u}_{33})_{LR}
&\simeq& 
\left[ 0.35 \tilde{A}_{a}^{u}+6.8 \times 10^{-4} \tilde{A}_{b}^{u}+3.0 \times 10^{-3}  \tilde{A}_{b'}^{u}
+1.7 \times 10^{-10} \tilde{A}_{c}^{u}
~\right] 
\left( \frac{500 ~\mbox{GeV}}{m_{\tilde{q}} }\right), \nn\\
(\delta^{u}_{12})_{LR}
&\simeq&-(\delta^{u}_{21})_{LR}\simeq 
7.4 \times 10^{-5} \left(\tilde{A}_{a}^{u}-\tilde{A}_{b}^{u}-
\tilde{A}_{b'}^{u}+\tilde{A}_{c}^{u}
\right)~
\left( \frac{500 ~\mbox{GeV}}{m_{\tilde{q}} }\right) ,\nn\\
(\delta^{u}_{13})_{LR}
&\simeq& 10^{-4}\left[
-7.9 \left(\tilde{A}_{a}^{u}-\tilde{A}_{b}^{u}\right)
-7.0 \times 10^{-2}\left(\tilde{A}_{b'}^{u}-\tilde{A}_{c}^{u}
\right)\right]~
\left( \frac{500 ~\mbox{GeV}}{m_{\tilde{q}} }\right), \nn\\
(\delta^{u}_{31})_{LR}
&\simeq& 10^{-3}\left[
-1.7 \left(\tilde{A}_{a}^{u}-\tilde{A}_{b'}^{u}\right)
-3.3 \times 10^{-3} \left(\tilde{A}_{b}^{u}-\tilde{A}_{c}^{u}
\right)\right]~
\left( \frac{500 ~\mbox{GeV}}{m_{\tilde{q}} }\right),
\label{LRu}\\
(\delta^{u}_{23})_{LR}
&\simeq&10^{-2}\left[
1.5 \left(\tilde{A}_{a}^{u}-\tilde{A}_{b}^{u}\right)
+1.3 \times 10^{-2} \tilde{A}_{b'}^{u}
+3.6 \times 10^{-5} \tilde{A}_{c}^{u}\right]~
\left( \frac{500 ~\mbox{GeV}}{m_{\tilde{q}} }\right) ,\nn\\
(\delta^{u}_{32})_{LR}
&\simeq&10^{-2}\left[
- 3.2 \left(\tilde{A}_{a}^{u}-\tilde{A}_{b'}^{u}\right)
-6.4 \times 10^{-3} \tilde{A}_{b}^{u}
-1.7 \times 10^{-5} \tilde{A}_{c}^{u}\right]~
\left( \frac{500 ~\mbox{GeV}}{m_{\tilde{q}} }\right).\nn
\ee

\vskip 0.5cm
\underline{\bf Down quark  sector ($LL$ and $RR$):}
%\vskip 0.5cm
\be
(\delta^d_{12})_{LL}
&=&  (\delta^d_{21})_{LL}\simeq 
1.2 \times 10^{-4}~\Delta a_L^{q}, \nn\\
(\delta^d_{13})_{LL}
&=& (\delta^d_{31})_{LL}
\simeq  7.8\times 10^{-3}~\Delta a_L^q, \nn\\
(\delta^d_{23})_{LL}
&=& (\delta^d_{32})_{LL}
\simeq  -1.5 \times 10^{-2}~\Delta a_L^{q},\nn\\
(\delta^d_{12})_{RR}
&=& (\delta^d_{21})_{RR}
\simeq -1.7 \times 10^{-1} ~\Delta a_R^{d}, \label{deltau1}
\\
(\delta^d_{13})_{RR}
&=& (\delta^d_{31})_{RR}
\simeq 1.4 \times 10^{-1}~\Delta a_R^{d}, \nn\\
(\delta^d_{23})_{RR}
&=& (\delta^d_{32})_{RR}
\simeq 4.7 \times 10^{-1}~\Delta a_R^{d}.\nn
\ee

\vskip 0.5cm
\underline{\bf Down quark  sector ($LR$): }
%\vskip 0.5cm
\be
(\delta^{d}_{11})_{LR}
&\simeq& 10^{-6}
\left[ 6.6  \left( \tilde{A}_{a}^{d}-\tilde{A}_{b}^{d}-\tilde{A}_{b'}^{d}+2 \tilde{A}_{c}^{d} \right)
~\right] ~\left( \frac{500 ~\mbox{GeV}}{m_{\tilde{q}} 
}\right), \nn\\
(\delta^{d}_{22})_{LR}
&\simeq& 10^{-5}
\left[ -4.1 \left( \tilde{A}_{a}^{d}-\tilde{A}_{b'}^{d}\right)+11\tilde{A}_{b}^{d}
+1.5\tilde{A}_{c}^{d}
~\right] ~\left( \frac{500 ~\mbox{GeV}}{m_{\tilde{q}} 
}\right), \nn\\
(\delta^{d}_{33})_{LR}
&\simeq&10^{-3} 
\left[ 2.3\tilde{A}_{a}^{d}+1.5\times 10^{-3}\tilde{A}_{b}^{d}+3.5\tilde{A}_{b'}^{d}
+1.4 \times 10^{-4}\tilde{A}_{c}^{d}
~\right] ~\left( \frac{500 ~\mbox{GeV}}{m_{\tilde{q}} 
}\right), \nn\\
 (\delta^{d}_{12})_{LR}
&\simeq&10^{-5} \left[~2.1
 \left( \tilde{A}_{a}^{d}-\tilde{A}_{b}^{d}-\tilde{A}_{b'}^{d}+ \tilde{A}_{c}^{d} \right)
\right]
~\left( \frac{500 ~\mbox{GeV}}{m_{\tilde{q}} 
}\right), \nn\\
(\delta^{d}_{21})_{LR}
&\simeq&10^{-5}\left[~
-1.3 \left( \tilde{A}_{a}^{d}-\tilde{A}_{b'}^{d} \right)
+3.4 \left( \tilde{A}_{b}^{d}-\tilde{A}_{c}^{d} \right)
\right]
\left( \frac{500 ~\mbox{GeV}}{m_{\tilde{q}} 
}\right),\\
\label{LRd}
(\delta^{d}_{13})_{LR}
&\simeq& 10^{-5}\left[
-1.8 \left( \tilde{A}_{a}^{d}-\tilde{A}_{b}^{d} \right)
-2.8 \left( \tilde{A}_{b'}^{d}-\tilde{A}_{c}^{d} \right)
\right]
\left( \frac{500 ~\mbox{GeV}}{m_{\tilde{q}} }\right), 
\nn\\
(\delta^{d}_{31})_{LR}
&\simeq& 10^{-4}
\left[
-8.4 \left( \tilde{A}_{a}^{d}-\tilde{A}_{b'}^{d} \right)
-10^{-3} \left( 5.6\tilde{A}_{b}^{d}-6.1\tilde{A}_{c}^{d} \right)
\right] 
 \left( \frac{500 ~\mbox{GeV}}{m_{\tilde{q}} }\right) ,\nn\\
(\delta^{d}_{23})_{LR}
&\simeq&10^{-5}\left[~
3.5 \tilde{A}_{a}^{d}-9.3 \tilde{A}_{b}^{d}+5.3 \tilde{A}_{b'}^{d}
+0.52 \tilde{A}_{c}^{d}
~\right] 
\left( \frac{500 ~\mbox{GeV}}{m_{\tilde{q}} }\right), \nn\\
(\delta^{d}_{32})_{LR}
&\simeq&10^{-3}\left[
-2.7 \left( \tilde{A}_{a}^{d}-\tilde{A}_{b'}^{d} \right)
-1.8 \times 10^{-3} \tilde{A}_{b}^{d}
-6.7 \times 10^{-5} \tilde{A}_{c}^{d}
\right] \left( \frac{500 ~\mbox{GeV}}{m_{\tilde{q}} }\right).\nn
\ee

\begin{table}[thb]
\begin{center}
\begin{tabular}{|c||c|c|}
 \hline
 &  Exp. bound  & $Q_6$ Model \\
%  & Exp. bound
   \hline  \hline

%\\ \hline
$\sqrt{|\mbox{Re}(\delta^d_{12})^2_{LL,RR}|}$
& $4.0 \times 10^{-2} ~\tilde{m}_{\tilde{q}}$
& $(LL)1.2 \times 10^{-4} \Delta a_L^q,(RR)1.7 \times 10^{-1}\Delta a_R^d$
\\ \hline
 $\sqrt{|\mbox{Re}(\delta^d_{12})_{LL}(\delta^d_{12})_{RR}|}$
& $2.8 \times 10^{-3} ~\tilde{m}_{\tilde{q}}$
&$4.5 \times 10^{-3}\sqrt{\Delta a_L^q \Delta a_R^d}$
\\ \hline
$\sqrt{|\mbox{Re}(\delta^d_{12})^2_{LR}|}$
 & $4.4 \times 10^{-3} ~\tilde{m}_{\tilde{q}}$
&$\sim 2 \times 10^{-5}(\tilde A^d_a-\tilde A^d_b-\tilde A^d_{b'}
+\tilde A^d_c)\tilde{m}_{\tilde{q}}^{-1}$
\\ \hline
 $\sqrt{|\mbox{Re}(\delta^d_{13})^2_{LL,RR}|}$
 & $9.8 \times 10^{-2} ~\tilde{m}_{\tilde{q}}$
&$(LL)7.8 \times 10^{-3}\Delta a_L^q,(RR)1.4 \times 10^{-1}\Delta a_R^d$
\\ \hline
$\sqrt{|\mbox{Re}(\delta^d_{13})_{LL}(\delta^d_{13})_{RR}|}$
&  $1.8 \times 10^{-2} ~\tilde{m}_{\tilde{q}}$
&$3.4 \times 10^{-2}\sqrt{\Delta a_L^q \Delta a_R^d}$
\\ \hline
 $\sqrt{|\mbox{Re}(\delta^d_{13})^2_{LR}|}$
& $3.3 \times 10^{-2} ~\tilde{m}_{\tilde{q}}$
&$\sim 2 \times 10^{-5}(\tilde A^d_a-\tilde A^d_b+\tilde A^d_{b'}
-\tilde A^d_c)\tilde{m}_{\tilde{q}}^{-1}$
\\ \hline
$\sqrt{|\mbox{Re}(\delta^u_{12})^2_{LL,RR}|}$
& $1.0 \times 10^{-1} ~\tilde{m}_{\tilde{q}}$
&$(LL)1.0 \times 10^{-4}\Delta a_L^q,(RR)4.5 \times 10^{-4}\Delta a_R^u$
\\ \hline
 $\sqrt{|\mbox{Re}(\delta^u_{12})_{LL}(\delta^u_{12})_{RR}|}$
& $1.7 \times 10^{-2} ~\tilde{m}_{\tilde{q}}$
&$2.1 \times 10^{-4}\sqrt{\Delta a_L^q \Delta a_R^u}$
\\ \hline
$\sqrt{|\mbox{Re}(\delta^u_{12})^2_{LR}|}$
& $3.1 \times 10^{-2} ~\tilde{m}_{\tilde{q}}$
&$\sim 7 \times 10^{-5}(\tilde A^u_a-\tilde A^u_b-\tilde A^u_{b'}
+\tilde A^u_c)\tilde{m}_{\tilde{q}}^{-1}$
 \\ \hline
$|(\delta^d_{23})_{LL,RR}|$
 & $8.2~ \tilde{m}_{\tilde{q}}^2$
 &$(LL)1.5 \times 10^{-2}\Delta a_L^q,(RR)4.7 \times 10^{-1}\Delta a_R^d$
 \\ \hline
 $|(\delta^d_{23})_{LR}|$
& $1.6 \times 10^{-2} ~\tilde{m}_{\tilde{q}}^2$
 &$\sim 5 \times 10^{-5}(\tilde A^d_a-\tilde A^d_b+\tilde A^d_{b'}
+0.1\tilde A^d_c)\tilde{m}_{\tilde{q}}^{-1}$
\\ \hline
\end{tabular}
\caption{Experimental bounds on  $\delta$'s and their theoretical values in $Q_6$ model,
 where the parameter
$\tilde{m}_{\tilde{q}}$ denotes
$m_{\tilde{q}} /500$ GeV, and $\Delta a_{L,R}$
 and $\tilde A$ are given 
in (\ref{deltaAt1}).}
\end{center}
\end{table}

\begin{table}[thb]
\begin{center}
\begin{tabular}{|c||c|c|}
 \hline
 &  Exp. bound  & $Q_6$ Model \\
%  & Exp. bound
   \hline  \hline

%\\ \hline
$|(\delta^e_{12})_{LL}|$
& $4.0 \times 10^{-5} ~\tilde{m}_{\tilde{\ell}}^2$
& $4.9 \times 10^{-3} \Delta a_L^{\ell} $
\\ \hline
 $|(\delta^e_{12})_{RR}|$
& $9 \times 10^{-4} ~\tilde{m}_{\tilde{\ell}}^2$
&$8.4 \times 10^{-8} \Delta a_R^{e} $
\\ \hline
$|(\delta^e_{12})_{LR}|$
 & $8.4 \times 10^{-7} ~\tilde{m}_{\tilde{\ell}}^2$
&$5.1 \times 10^{-6}(\tilde A_{b'}^e-\tilde A_c^e)\tilde{m}_{\tilde{\ell}}^{-1}$
\\ \hline
%$|(\delta^\ell_{21})_{LR}|$
% & $\bf ??8.4 \times 10^{-7} ~\tilde{m}_{\tilde{\ell}}^2$
%&$2.5 \times 10^{-8}(\tilde A_2^\ell-\tilde A_4^\ell)\tilde{m}_{\tilde{\ell}}^{-1}$
%\\ \hline
 $|(\delta^e_{13})_{LL}|$
 & $2 \times 10^{-2} ~\tilde{m}_{\tilde{\ell}}^2$
&$1.7 \times 10^{-5}\Delta a_L^{\ell} $
\\ \hline
$|(\delta^e_{13})_{RR}|$
&  $3 \times 10^{-1} ~\tilde{m}_{\tilde{\ell}}^2$
&$5.9 \times 10^{-2}\Delta a_R^{e} $
\\ \hline
 $|(\delta^e_{13})_{LR}|$
& $1.7 \times 10^{-2} ~\tilde{m}_{\tilde{\ell}}^2$
&$3.1 \times 10^{-7}(\tilde A_b^e-\tilde A_{b'}^e)\tilde{m}_{\tilde{\ell}}^{-1}$
\\ \hline
% $|(\delta^\ell_{31})_{LR}|$
%& $\bf ??1.7 \times 10^{-2} ~\tilde{m}_{\tilde{\ell}}^2$
%&$1.1 \times 10^{-3}(\tilde A_2^\ell-\tilde A_5^\ell)\tilde{m}_{\tilde{\ell}}^{-1}$
%\\ \hline
$|(\delta^e_{23})_{LL}|$
& $2 \times 10^{-2} ~\tilde{m}_{\tilde{\ell}}^2$
&$8.4 \times 10^{-8}\Delta a_L^{\ell} $
\\ \hline
 $|(\delta^e_{23})_{RR}|$
& $3 \times 10^{-1} ~\tilde{m}_{\tilde{\ell}}^2$
&$1.4 \times 10^{-6} \Delta a_R^{e} $
\\ \hline
$|(\delta^e_{23})_{LR}|$
& $1 \times 10^{-2} ~\tilde{m}_{\tilde{\ell}}^2$
&$1.5 \times 10^{-9}(\tilde A_b^e-\tilde A_{b'}^e)\tilde{m}_{\tilde{\ell}}^{-1}$
 \\ \hline
% $|(\delta^\ell_{32})_{LR}|$
%& $\bf ?? 1 \times 10^{-2} ~\tilde{m}_{\tilde{\ell}}^2$
%&$2.5 \times 10^{-8}(\tilde A_2^\ell-\tilde A_5^\ell)\tilde{m}_{\tilde{\ell}}^{-1}$
 %\\ \hline
 $~~|(\delta^e_{23})_{LL}(\delta^e_{13})_{LL}|~~$
& $1 \times 10^{-4} ~\tilde{m}_{\tilde{\ell}}^2$
 &$1.4 \times 10^{-12} (\Delta a_L^{\ell})^2 $
\\ \hline
 $|(\delta^e_{23})_{RR}(\delta^e_{13})_{RR}|$
& $9 \times 10^{-4} ~\tilde{m}_{\tilde{\ell}}^2$
 &$8.4 \times 10^{-8}  (\Delta a_R^{e})^2$
\\ \hline
 $|(\delta^e_{23})_{LL}(\delta^e_{13})_{RR}|$
& $2 \times 10^{-5} ~\tilde{m}_{\tilde{\ell}}^2$
 &$5.0 \times 10^{-9}  \Delta a_L^{\ell} \Delta a_R^{e}$
\\ \hline
 $|(\delta^e_{23})_{RR}(\delta^e_{13})_{LL}|$
& $2 \times 10^{-5} ~\tilde{m}_{\tilde{\ell}}^2$
 &$2.4 \times 10^{-11}  \Delta a_L^{\ell} \Delta a_R^{e} $
\\ \hline
\end{tabular}
\caption{Experimental bounds on  $\delta$'s and the theoretical values in $Q_6$ model, where
the parameter
$\tilde{m}_{\tilde{\ell}}$ denote
$m_{\tilde{\ell}} /100$ GeV and $\Delta a_{L,R}$ and $\tilde A$ are given 
in (\ref{deltaAt1}).}
\end{center}
\end{table}
In Table II and III, 
theoretical values of certain $\delta$'s
calculated  above and 
their experimental bounds are summarized. 
We see from the tables that to satisfy the experimental 
constraints, 
the SSB parameters
$\Delta a_R^d,~\Delta a_L^{\ell},
~\tilde A_{b'}^e$ and $\tilde A_c^e$
 of the present $Q_6$ model should satisfy
\be
\Delta a_R^d<10^{-1},~~\Delta a_L^{\ell}<10^{-2},
~~\tilde A_{b'}^e-\tilde A_c^e<10^{-1},
\label{FCNCcondition}
\ee 
while the other SSB parameters are allowed to  be of
$O(1)$.
So, one can fairly say that
 $Q_6$ symmetry can soften the SUSY flavor problem. 
Note also that the degree of degeneracy 
of the left-handed squark masses $\Delta a_L^q$ 
does not need to be very accurate. We find that 
because of the constraint on $\Delta a_R^d$
given in (\ref{FCNCcondition}),
 $\Delta a_L^q<10^{+1}$ is sufficient to satisfy
 all the constraints. This has an important consequence
 for  proton decay as we will see in the next section.

\section{Proton Decay}
In this section we consider proton decay,  which is a process
reflecting the flavor structure of a model.
\subsection{$Q_6$ invariant  baryon and lepton number 
violating operators }
 In supersymmetric models, the  baryon and/or lepton number 
violating operators of lower dimensions are \cite{protondecay}: 
\begin{enumerate}
\item dimension-four R parity violating operators, and
\item dimension-five baryon and lepton number violating operators.
\end{enumerate}   
Both operators are controlled by the flavor structure of a model. 
As for the dimension-four R parity violating operators, 
$Q_6$ flavor symmetry in the model considered
in the previous sections allows only lepton number 
violating operators:
\be
LLE^c~&:&~\lambda_{113}L_I L_I E^c_3 +\lambda_{311}L_3 L_I E^c_I
+\lambda_{333}L_3 L_3 E^c_3
+\lambda f_{IJK}L_I L_J E^c_K, \nn \\
LQD^c~&:&~\lambda'_{132}L_I \left(i \sigma^2 \right)_{IJ}Q_3 D^c_J
+\lambda'_{123}L_I \left( \sigma^1 \right)_{IJ}Q_J D^c_3, \nn \\
H^dNH^u~&:&~\tilde{\lambda}_{312}H^d_3N_IH_I^u+
\tilde{\lambda}_{333}H^d_3N_3H^u_3
+\tilde{\lambda}f_{IJK}H_I^dN_JH_K^u ,\nn \\
NNN~&:&~\tilde{\lambda}'f_{IJK}N_IN_JN_K,
\ee
and those in which  $L_I$ is replaced by $H^d_I$. 
Note that the baryon number violating operator 
$U^c D^c D^c$ is forbidden by $Q_6$ symmetry. 
Therefore,  
the dimension-four  operators in the present model can not 
mediate proton decay.

We next look at dimension-five operators. 
The baryon number violating  dimension-five operators (which are allowed by 
R parity and $B-L$ symmetry) can be  
written as \cite{protondecay,rudaz,raby,yanagida}
\be
W_5=\frac{1}{M}\sum_{i,j,k,l=1 \sim 3}\left[ \frac 12 C_L^{ijkl}Q_i Q_j Q_k L_l+C_R^{ijkl}
U^c_i E^c_j U^c_k D^c_l \right],
\label{w51}
\ee
where the first and the second term is called 
the LLLL and RRRR operator, respectively.
In grand unified theories(GUTs), effective dimension-five operators 
can be generated by integrating out 
colored Higgs multiplets \cite{protondecay,rudaz}, and 
therefore the size of the coefficients
of the operators strongly depends 
the Yukawa matrices. 
For the minimal SUSY $SU(5)$ GUT \cite{yanagida}, one obtains
\be
M=M_{H_C},~C_L^{ijkl}=C_R^{ijkl}=Y_U^{ij}Y_D^{kl}
~~{\mbox{at }}M_{GUT},
\ee 
where $M_{H_C}$ is the colored Higgs mass of  order of the GUT scale, 
and $Y_U,~Y_D$ are 
Yukawa coupling matrices appearing in the superpotential
\be
W_Y=\frac 14 Y_U^{ij}{\bf 10}^i {\bf 10}^j H+\sqrt{2}Y_D^{ij}
{\bf 10}^i \bar{\bf 5}^j \bar H. 
\ee 
Unfortunately, the minimal SUSY $SU(5)$ GUT 
should be excluded by the decay mode 
$p \to K^+ \bar \nu$ if the  gauge coupling unification
should be strictly satisfied \cite{pierce}.

If we do not assume  any GUTs, 
 the baryon number violating operators could be generated by 
some unknown Planck scale physics.
In this case, the mass parameter $M$ in (\ref{w51}) is given by
\be
M=M_{PL}=2 \times 10^{18}\mbox{GeV},
\ee
while the coefficients $C_{L,R}^{ijkl}$ remain undetermined.
So,  the operators 
are supplied with  a  suppression factor $1/{M_{PL}}$ which should be
compared with  $1/{M_{H_C}}\simeq 1/10^{16}~\mbox{GeV}^{-1}$ 
in the  GUT case. However, 
this suppression is not sufficient to keep
the proton stable, unless that 
the coefficients $C$'s are smaller than
$\sim O(10^{-7})$.
An efficient tool to suppress or to forbid   proton decay is 
symmetry.
Many authors 
\cite{murayama,benhamo,carone} 
have proposed a model in which both
the gross structure of the baryon and lepton number violating  operators
and the  structure of the Yukawa couplings
are fixed by a single flavor symmetry and  its breaking.
If this is realized, the flavor symmetry can be tested by  proton decay, too.
In  the models 
of \cite{murayama,benhamo,carone}, a certain set of
flavon fields is needed to form invariants, and
the flavor symmetry is assumed to be broken at a superhigh energy scale.
In contrast to these models, $Q_6$ flavor symmetry
is broken (spontaneously and
at most softly) only at a low energy scale which is comparable with the SUSY
breaking scale. So, it is natural to assume  that $Q_6$ flavor symmetry is 
intact at the Planck scale, too;
we do not have to introduce
flavon fields. Moreover, $Q_6$ is
nonabelian, which can indeed reduce the number of independent
coefficients drastically, as we will see now.
We find that the relevant superpotential
containing the baryon number 
violating $Q_6$ invariant 
dimension-five operators generated at the Planck scale 
can be written as
\be
W_5^{Q_6}=\frac{1}{M_{PL}}\sum_{I,J=1,2}\left[ C_L Q_I Q_I Q_3 L_3+C_R^{(1)}E_I^c 
\left(i \sigma^2 \right)_{IJ} U_J^c 
U_3^c D_3^c +C_R^{(2)}E_I^c 
\left( i \sigma^2 \right)_{IJ}U_1^c U_2^c D_J^c \right],
\label{dim5q6}
\ee
where the superfields in (\ref{dim5q6}) are in the flavor eigenstates.
To obtain Eq.(\ref{dim5q6}), we  have not
assumed any symmetries such as R-parity and $B-L$ except for $Q_6$ symmetry, where the $Q_6$ assignment is given in Table I.

\subsection{Gross structure of the dimension-five operators
and the lowest order approximation}
\begin{figure}[ht]
\unitlength=1mm
\begin{picture}(100,40)

\includegraphics[width=5cm]{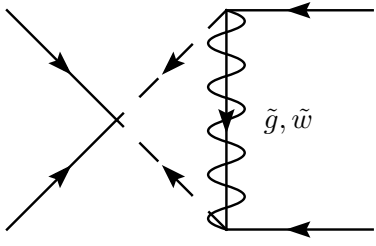}

\put(-15,14){\bf $\tilde{g},\tilde{w}$}

\end{picture}
\caption{\label{graph1}
One-loop diagrams contributing to the 
the effective four-fermi Lagrangian (\ref{effectiveL}).
We consider only gluino ($\tilde{g}$) and wino ($\tilde{w}$)
dressings in the lowest order approximation.
}
\end{figure}

As we can see from (\ref{dim5q6}), $Q_6$ allows only three
independent coefficients, 
$ C_L, C_R^{(1)}$ and $C_R^{(2)}$.
We will see moreover that the first term, the LLLL operator,
gives the most dominant contribution to  proton decay,
while the RRRR operators can be neglected
in the lowest order approximation.
Consequently, the relative size of all the partial decay rates is 
fixed in this approximation,  once the SSB sector is fixed.
To begin with, we recall two basic facts:
\begin{enumerate}
\item
Since the operators in the first two terms in $ W_5^{Q_6}$ contain 
quark fields of the third generation in the flavor eigenstate, 
small mixing parameters appear when fields are 
rewritten in terms of
the mass eigenstates, that is, 
\be
\Phi^f_3=V_{3I}\Phi^m_I,~~I=1,2,
\ee
where the subscripts $f$ and $m$ denote 
the flavor and mass eigenstates, respectively. 
The mixing parameters $V_{3I}$ will be multiplied with
the coefficients
$C$'s in the decay amplitudes, so that the condition 
on  $C$'s can be relaxed if $V_{3I}$ are small (see (\ref{UuL})$\sim$(\ref{UdR})).
Note that within the framework of the MSSM, there is no such suppression.

\item
In most of models, 
the  universality of the SSB parameters at the GUT or Planck scale 
is assumed to suppress 
the SUSY contributions to FCNCs.
This assumption has an important consequence that 
the gluino  contributions to proton decay are negligibly 
small if all the squark masses are degenerate \cite{belyaev}.
(In Fig. \ref{graph1}. we show a gaugino dressing diagram which
contributes to an effective  four-fermion operator.)
However, in our case, we do not 
assume the  universality.
We have seen in the previous section that 
the degeneracy of the squark masses of the first two
generations is almost exact due to $Q_6$ symmetry.
Moreover, 
to suppress  the SUSY contributions to FCNCs
the degeneracy of the
first two and third generations 
needs not to be accurate (see Eq. (\ref{FCNCcondition}) and the
discussions below).
This means
 that the gluino dressing diagrams may not
be negligible \cite{chadha} in our case; we will have 
to investigate it.
\end{enumerate}

We now argue that the 
first term, the LLLL operator, is  the most dominant one.
It is known that for the RRRR operators, the dominant contributions 
come from the gluino dressing diagrams.  From the 
superpotential (\ref{dim5q6}), we first observe
that
the second term in (\ref{dim5q6}), the first RRRR operator,
 contains two quark fields of the 
 third generation, implying that two small mixing parameters 
 will be multiplied
when going to the mass eigenstates.
Further, the third term, the second RRRR term,  contains only  quark fields
 of the first two generation.
That is, the gluino dressing contributions 
 vanish because the squark masses of 
the first two generations are almost degenerate  thanks to $Q_6$ symmetry.
Thus, the LLLL operator is the only one 
which should be considered in the lowest order approximation, as we will do
it in the following discussions.

Note that the LLLL operator
 contains two third generation fields $Q_3$ and $L_3$, 
but the $(3,1)$ element of the mixing matrix
 $U_{eL}$ (see (\ref{UeL})) is equal to one, so that
 it does not act as a suppression factor.
The dominant diagrams for the LLLL operator are those
with gluino dressing.
The zino and photino dressing 
diagrams have the same structure as the gluino ones, 
but they are negligibly small because the corresponding gauge couplings are small. 
So we ignore them in our calculations.
We have calculated the higgsino dressing contributions 
and found that they can be neglected, too,  if $\tan \beta<10$. 
So we assume this to simplify our calculations.
We may further approximate that the squark masses are diagonal
in the super CKM basis. To see this, we recall that the squark
mass squared in the super CKM basis can be written as
\be
{\bf m}^2_a =({\bf m}_{f_a})^2+\left( \begin{array}{ccc} 
\Delta_{LL}^a & \Delta_{LR}^{a*}\\
\Delta_{LR}^{aT} & \Delta_{RR}^{aT}\\
\end{array}\right),~~a=u,d,e
\ee
where ${\bf m}_{f_a}$ is the 
diagonal fermion mass matrix of the flavor $a$, and $\Delta$'s are given in (\ref{Delta1}).  
As we can see from (\ref{LRell}),(\ref{LRu}) and (\ref{LRd}), the nondiagonal elements are 
sufficiently small.        

The effective four-fermi Lagrangian
${\cal L}_{\mbox{eff}}$ can be obtained from the diagrams
shown in Fig. \ref{graph1} , where as argued we consider only gluino and wino
dressings. We find \footnote{We use the notation 
of \cite{arafune,goto}.}
\be
{\cal L}_{\mbox{eff}}=\frac{1}{(4 \pi)^2 M_{PL}}\left[ 
\sum_{M=d,s \atop l = e,\mu}C_{LL}(udue)^{1M1l}(ud^M)(ue^{l})
+\sum_{M,N=d,s \atop l = e,\mu, \tau}C_{LL}(udd\nu)^{1MNl}(ud^M)
(d^N \nu^{l})\right],
\label{effectiveL}
\ee

\be
C_{LL}(udue)^{1M1l}&=&{\tilde C}^{\tilde g}_{LL}(udue)^{1M1l}
+{\tilde C}^{\tilde w}_{LL}(udue)^{1M1l},  \nn \\
C_{LL}(udd\nu)^{1MNl}&=&{\tilde C}^{\tilde g}_{LL}(udd\nu)^{1MNl}
+{\tilde C}^{\tilde w}_{LL}(udd\nu)^{1MNl},
\ee
where $\tilde C^{\tilde g}$ and $\tilde C^{\tilde w}$ stand for the contributions
coming  from 
the gluino and wino dressing diagrams, respectively. 
These are explicitly calculated to be 
\be
{\tilde C}^{\tilde g}_{LL}(udd\nu)^{1MNl}&=&4 \pi \frac{4 \alpha_3}{3} 
\sum_{I=1,2} U_{uL}^{I1}U_{dL}^{IM}U_{dL}^{3N}U_{eL}^{3 l} 
\left( F^{\tilde g}(a_L^q,b_L^q)-F^{\tilde g}
(a_L^q,a_L^q) \right), \nn \\
{\tilde C}^{\tilde w}_{LL}(udd\nu)^{1MNl}&=&4 \pi \alpha_2 
\sum_{I=1,2}\left[ U_{uL}^{I1}U_{dL}^{IM}U_{dL}^{3N}U_{eL}^{3 l} 
\left( F^{\tilde w}(a_L^q,a_L^q)+F^{\tilde w}
(b_L^q,b_L^{\ell})\right) \right. , \nn \\ 
&~&-\left. U_{uL}^{31}U_{dL}^{IM}U_{dL}^{IN}U_{eL}^{3 l}
\left( F^{\tilde w}(a_L^q,b_L^q)+F^{\tilde w}
(a_L^q,b_L^{\ell})\right)\right],
\label{uddnu}
\ee

\be
{\tilde C}^{\tilde g}_{LL}(udue)^{1M1l}&=&-4 \pi \frac{4 \alpha_3}{3} 
\sum_{I=1,2} U_{uL}^{I1}U_{dL}^{IM}U_{uL}^{31}U_{eL}^{3 l} 
\left( F^{\tilde g}(a_L^q,b_L^q)-F^{\tilde g}
(a_L^q,a_L^q) \right), \nn \\
{\tilde C}^{\tilde w}_{LL}(udue)^{1M1l}&=&-4 \pi \alpha_2 
\sum_{I=1,2}\left[ U_{uL}^{I1}U_{dL}^{IM}U_{uL}^{31}U_{eL}^{3 l} 
\left( F^{\tilde w}(a_L^q,a_L^q)+F^{\tilde w}
(b_L^q,b_L^{\ell})\right) \right. ,\nn \\
&~&-\left. U_{uL}^{I1}U_{dL}^{3M}U_{uL}^{I1}U_{eL}^{3 l}
\left( F^{\tilde w}(a_L^q,b_L^q)+F^{\tilde w}(a_L^q,b_L^{\ell})\right)\right],
\label{udue}
\ee
where the loop functions are defined as \cite{arafune,goto}
\be
F^{\tilde g}(a_L^q,b_L^q)&=&\frac{1}{m_{\tilde g}} \frac{1}{x_{g1}-x_{g3}}
\left[ \frac{x_{g1}\ln x_{g1}}{x_{g1}-1}-\frac{x_{g3}\ln x_{g3}}{x_{g3}-1}\right], \nn \\
F^{\tilde w}(a_L^q,b_L^{\ell})&=&\frac{1}{m_{\tilde w}} \frac{1}{x_{w1}-y_{w3}}
\left[ \frac{x_{w1}\ln x_{w1}}{x_{w1}-1}-\frac{y_{w3}\ln y_{w3}}{y_{w3}-1}\right], \nn
\ee
with
\be
x_{g,w1}&=&\frac{m^2_{\tilde q}}{m^2_{\tilde g,\tilde w}}a_L^q,~
x_{g,w3}=\frac{m^2_{\tilde q}}{m^2_{\tilde g,\tilde w}}b_L^q,~
y_{w3}=\frac{m^2_{\tilde \ell}}{m^2_{\tilde w}}b_L^{\ell}.~
\label{loopfn}
\ee
We also use $
\Delta_Q \equiv \Delta a_L^q=a_L^q-b_L^q$
(which is first introduced in (\ref{deltaAt1})) and 
\be
r_g &=&
\frac{m^2_{\tilde q}}{m^2_{\tilde g}}.
\ee
Then we calculate the decay amplitudes
 as a function of $x_{g3}$ for given values of 
$r_g \Delta_Q$ and $y_{w3}$,
while for simplicity  we  assume the GUT relation between the wino and gluino mass
\footnote{If we change the GUT relation (\ref{gutr1}),
the results below can change only slightly.} 
\be
m_{\tilde w}=0.27 m_{\tilde g}.
\label{gutr1}
\ee
The unitary matrices $U_{(u,d,e)L}$ in $C_{LL}$ are explicitly given in 
(\ref{UuL}), (\ref{UdL}) and (\ref{UeL}), where
the individual phases $\theta^{u,d}$ and 
$\theta_3^{u,d}$ in the quark sector are not fixed
(see(\ref{parameters})). However, 
it is found from (\ref{UuL}) and (\ref{UdL}) that the phase 
dependence of the combinations 
appearing in $C_{LL}$, that is, $U_{uL}^{I1}U_{dL}^{I(1,2)}$ and
$U_{(u,d)L}^{I1}U_{(u,d)L}^{I1}$, is small,
and moreover the absolute size of the suppression 
factor $U_{(u,d)L}^{3I}$ is independent of the phases. 
Therefore, in the following calculations we choose 
\be
\theta^u=\theta^u_3=0,~\theta^d_3-\theta^d=-1.25
\label{phasechoice}
\ee
without significantly changing the results.

\subsection{Gluino versus wino  contributions}
\begin{figure}[!t]
  \centering
  \hspace*{-5mm}
  \includegraphics[width=8cm]{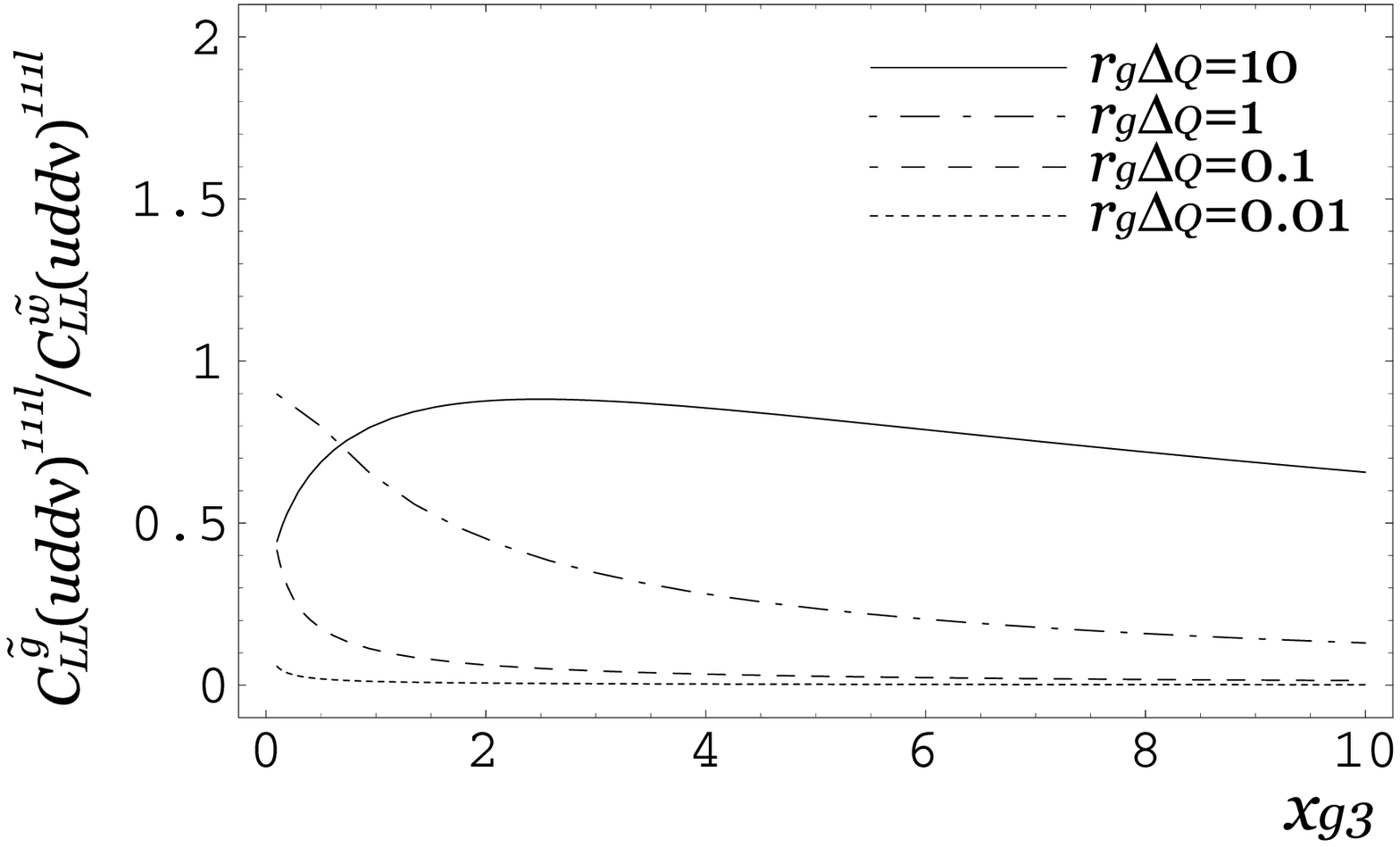}
  \hspace{5mm}
  \includegraphics[width=8cm]{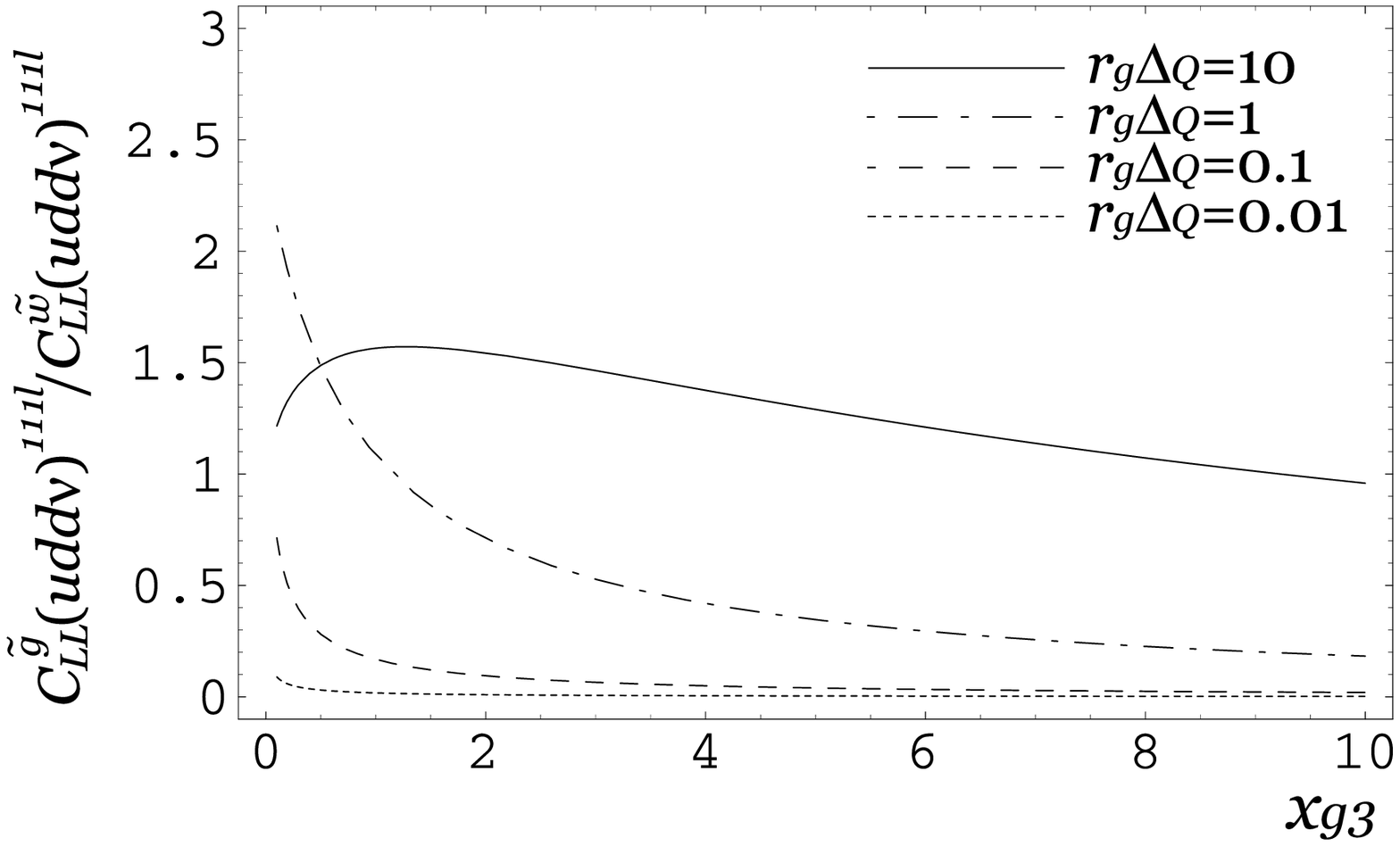} \\
\vspace{1cm}
  \hspace*{-5mm}
  \includegraphics[width=8cm]{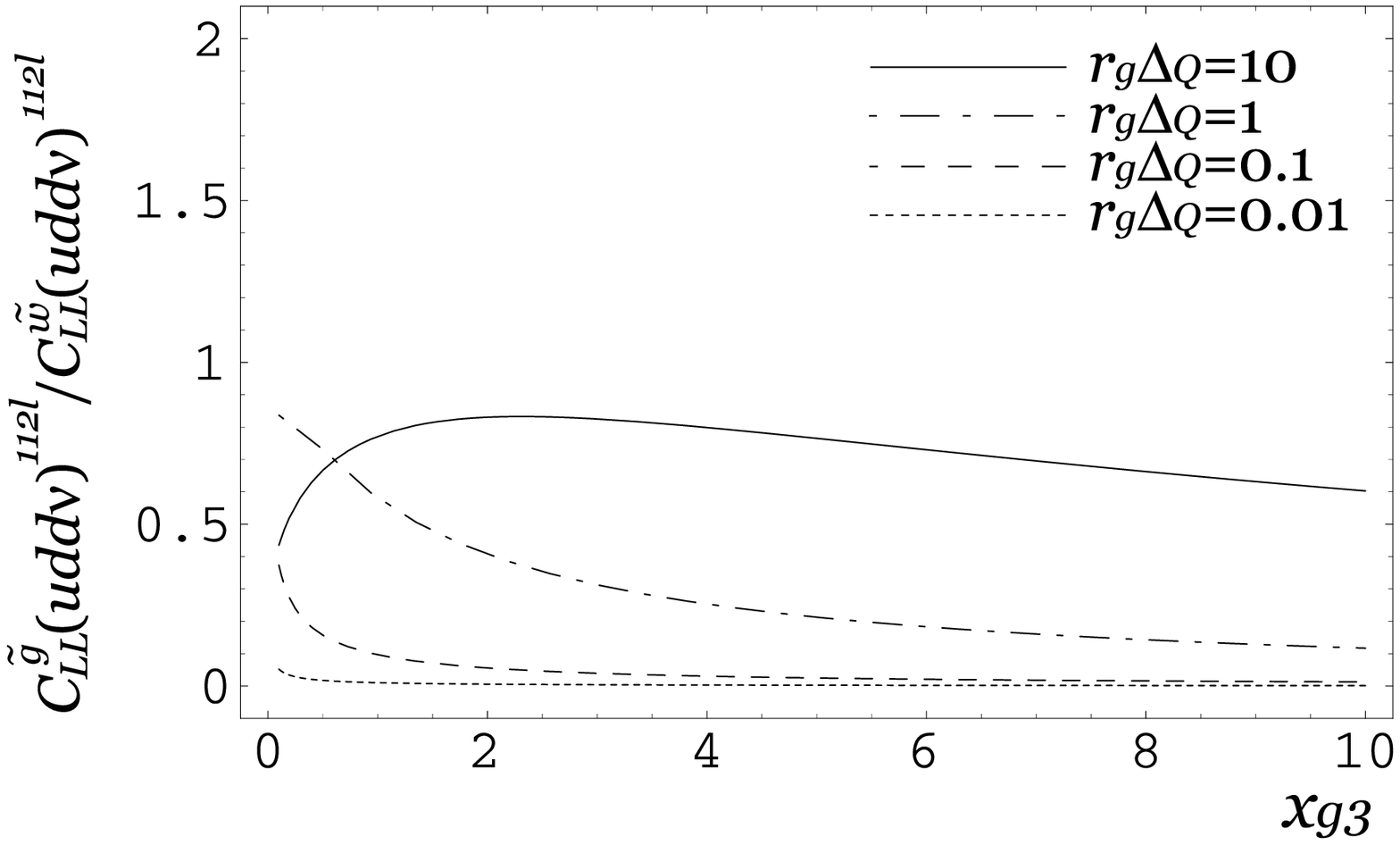}
  \hspace{5mm}
  \includegraphics[width=8cm]{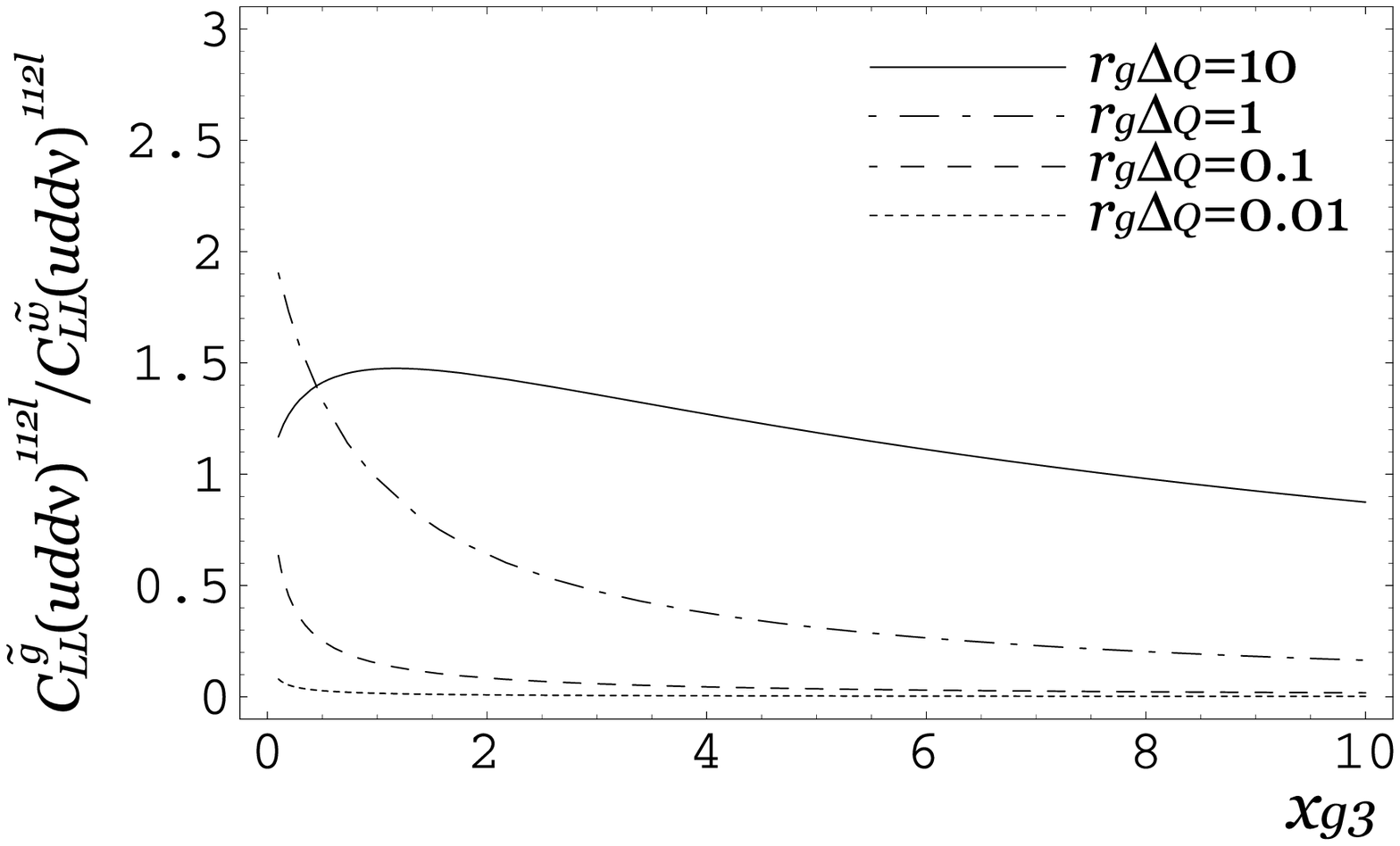} \\
  \vspace{1cm}
  \hspace*{-5mm}
  \includegraphics[width=8cm]{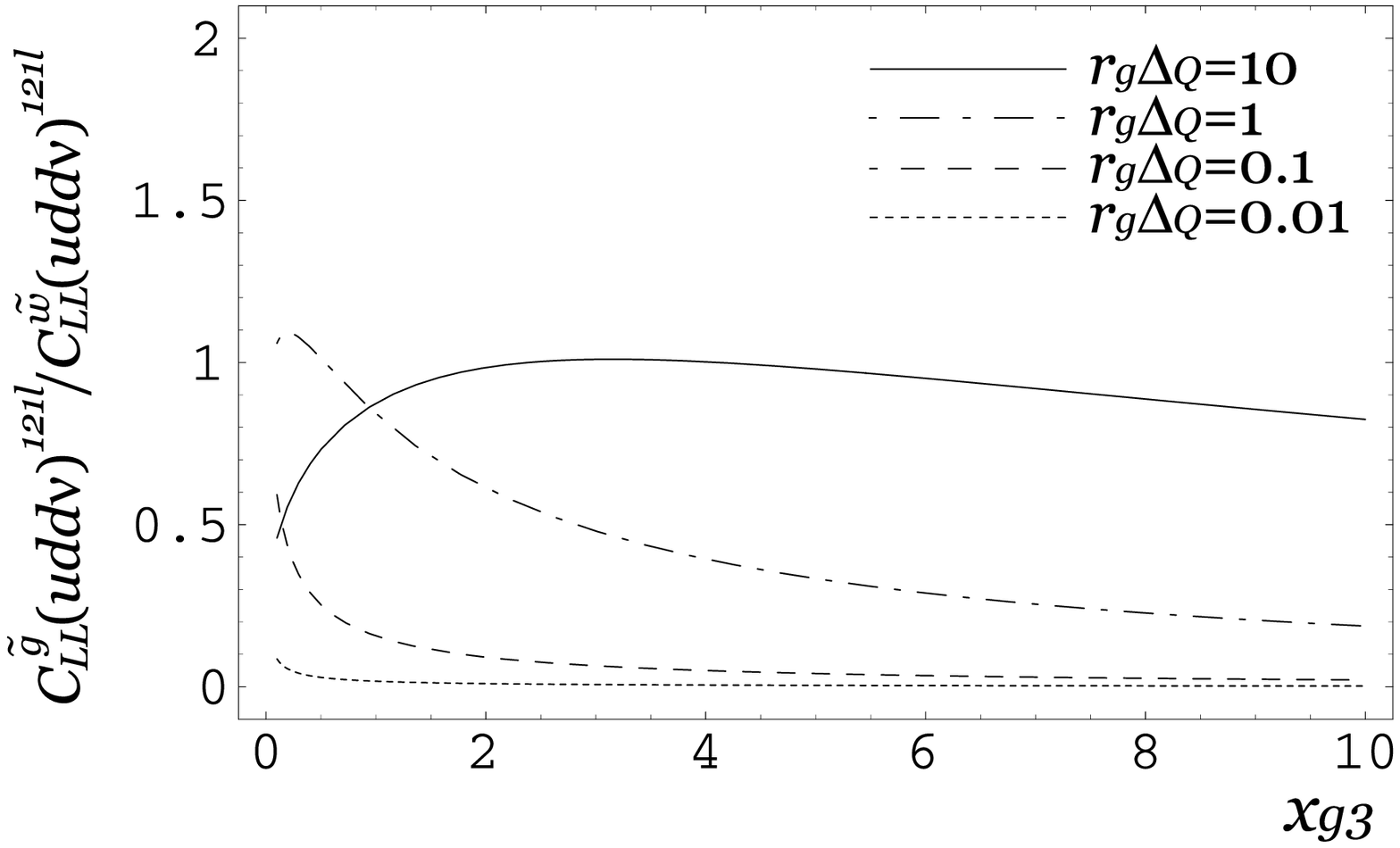}
  \hspace{5mm}
  \includegraphics[width=8cm]{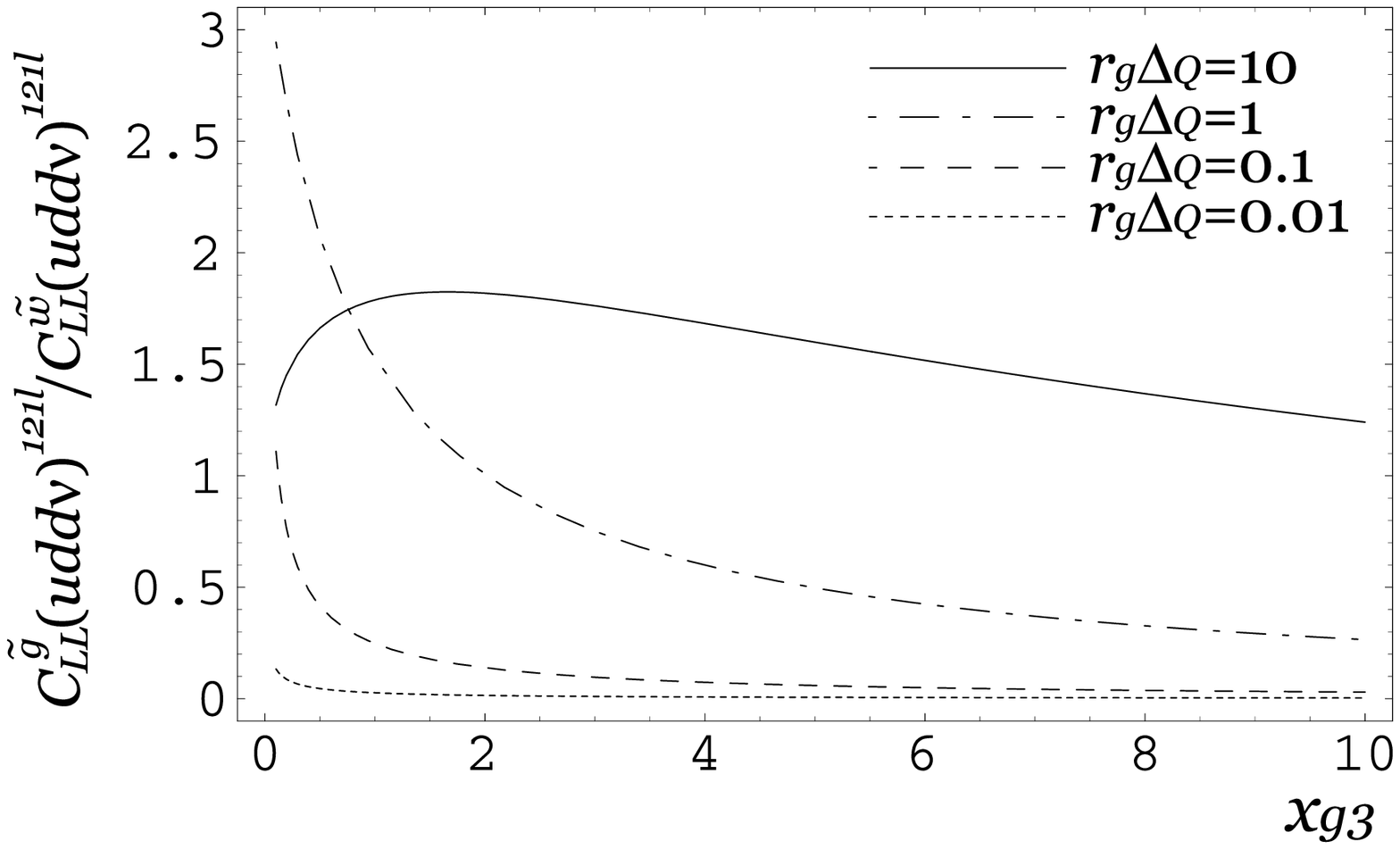} \\
  \caption{The ratio $C^{\tilde g}(udd\nu)/C^{\tilde w}(udd\nu)$ is plotted
  as a function of $x_{g3}$ for
$r_g\Delta_Q=(10,~1,~0.1,~0.01)$, which expresses
  the degree of the degeneracy of the squark masses, where
$x_{g3}$ is defined in (\ref{loopfn}).
$C^{\tilde g}$'s stand for the gluino contributions
and $C^{\tilde w}$'s  for the  wino ones, respectively.
The figures in the left column correspond to $y_{w3}=1$, 
  and those in the right column to $y_{w3}=10$.
The corresponding amplitudes are responsible for the anti-neutrino modes.}
  \label{dressingratio1}
\end{figure}

\begin{figure}[!t]
  \centering
  \hspace*{-5mm}
  \includegraphics[width=8cm]{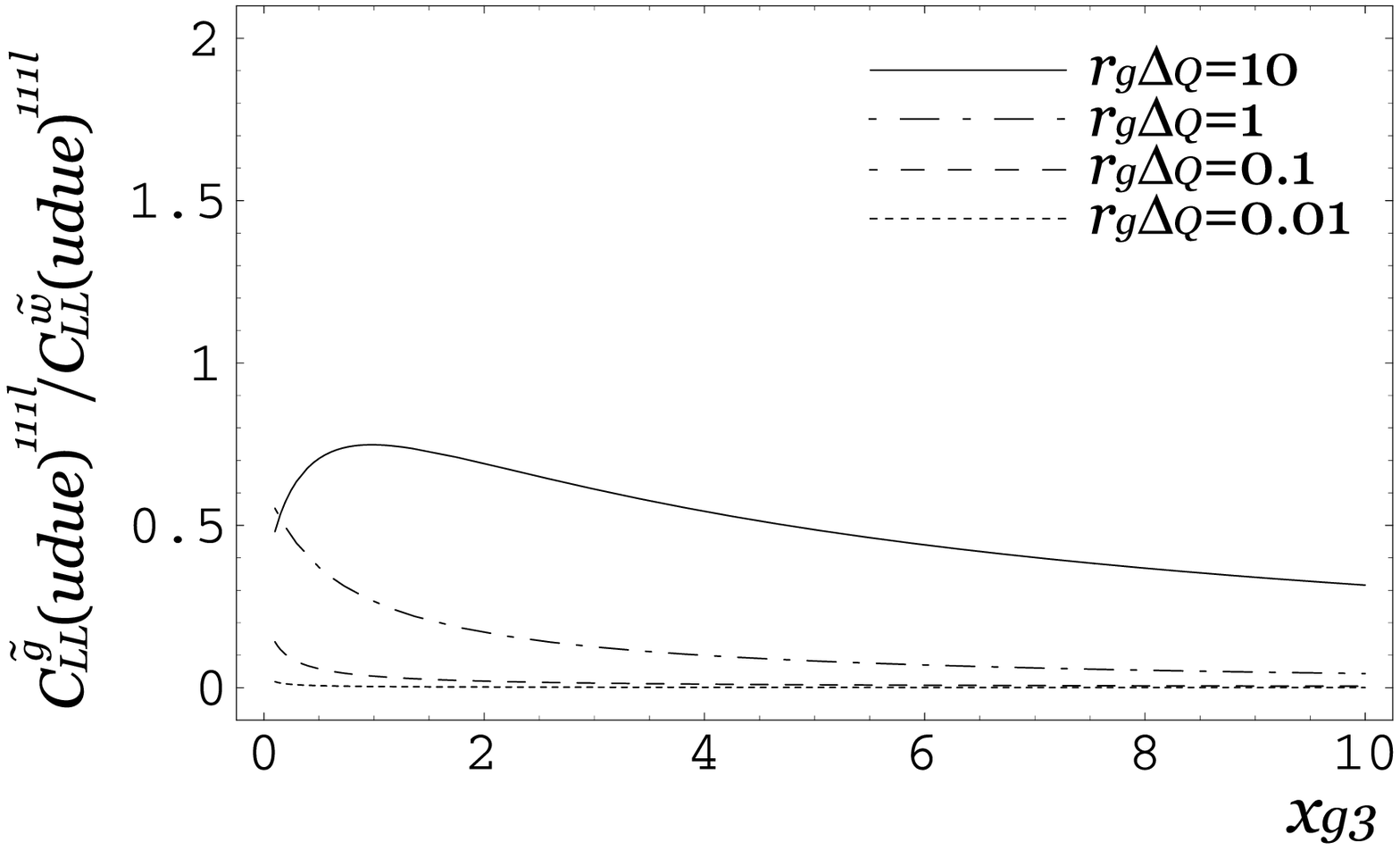}
  \hspace{5mm}
  \includegraphics[width=8cm]{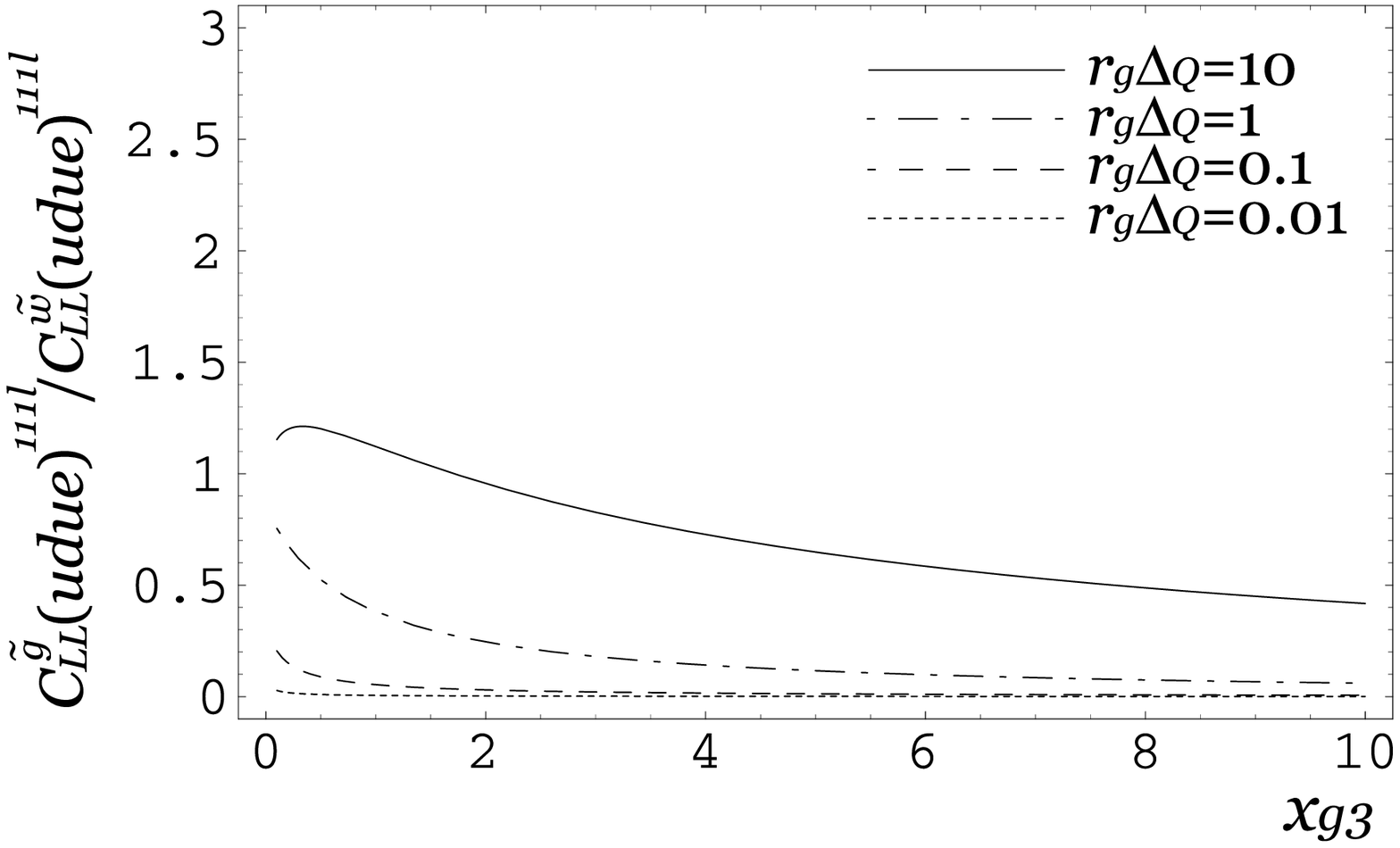} \\
\vspace{1cm}
  \hspace*{-5mm}
  \includegraphics[width=8cm]{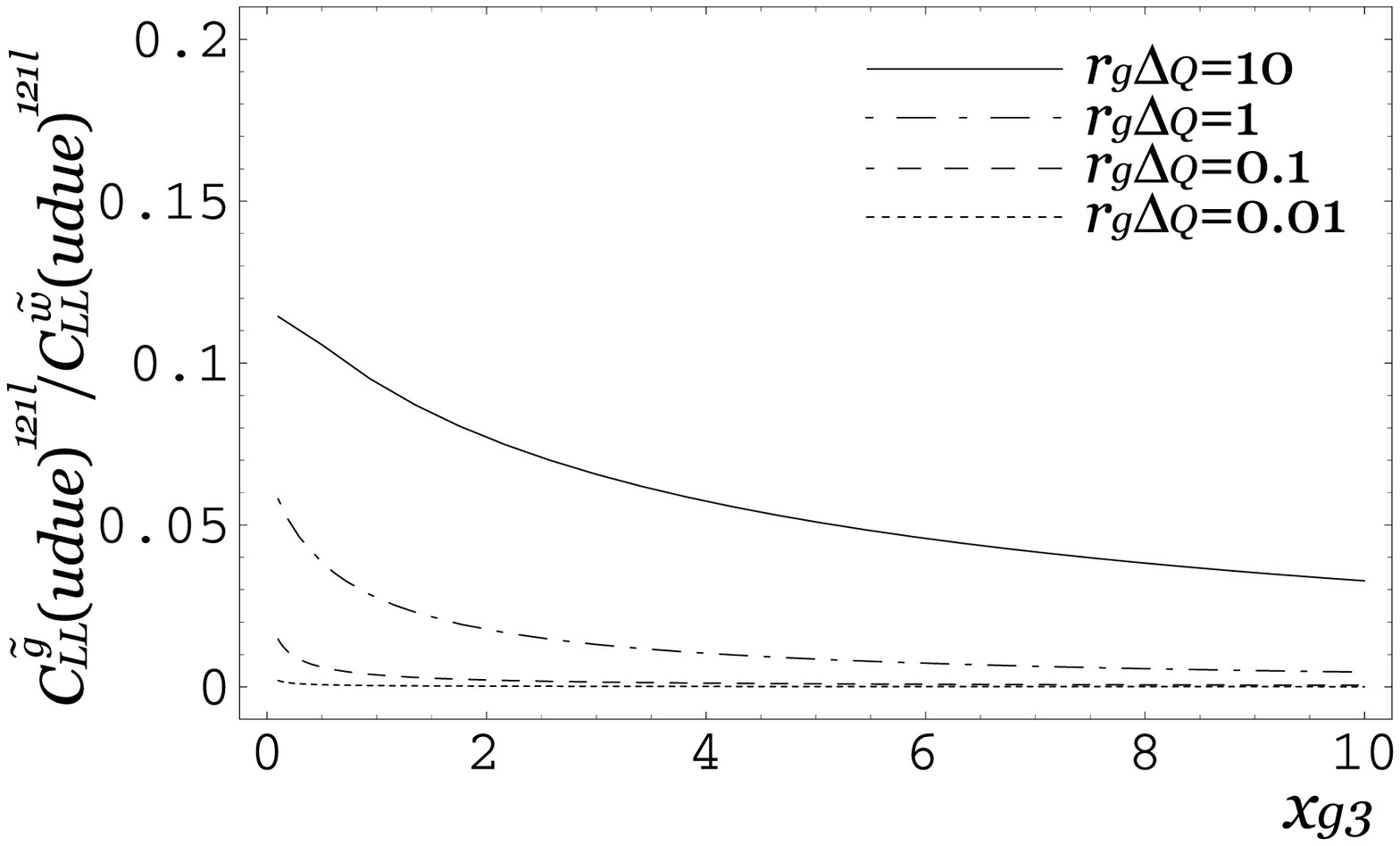}
  \hspace{5mm}
  \includegraphics[width=8cm]{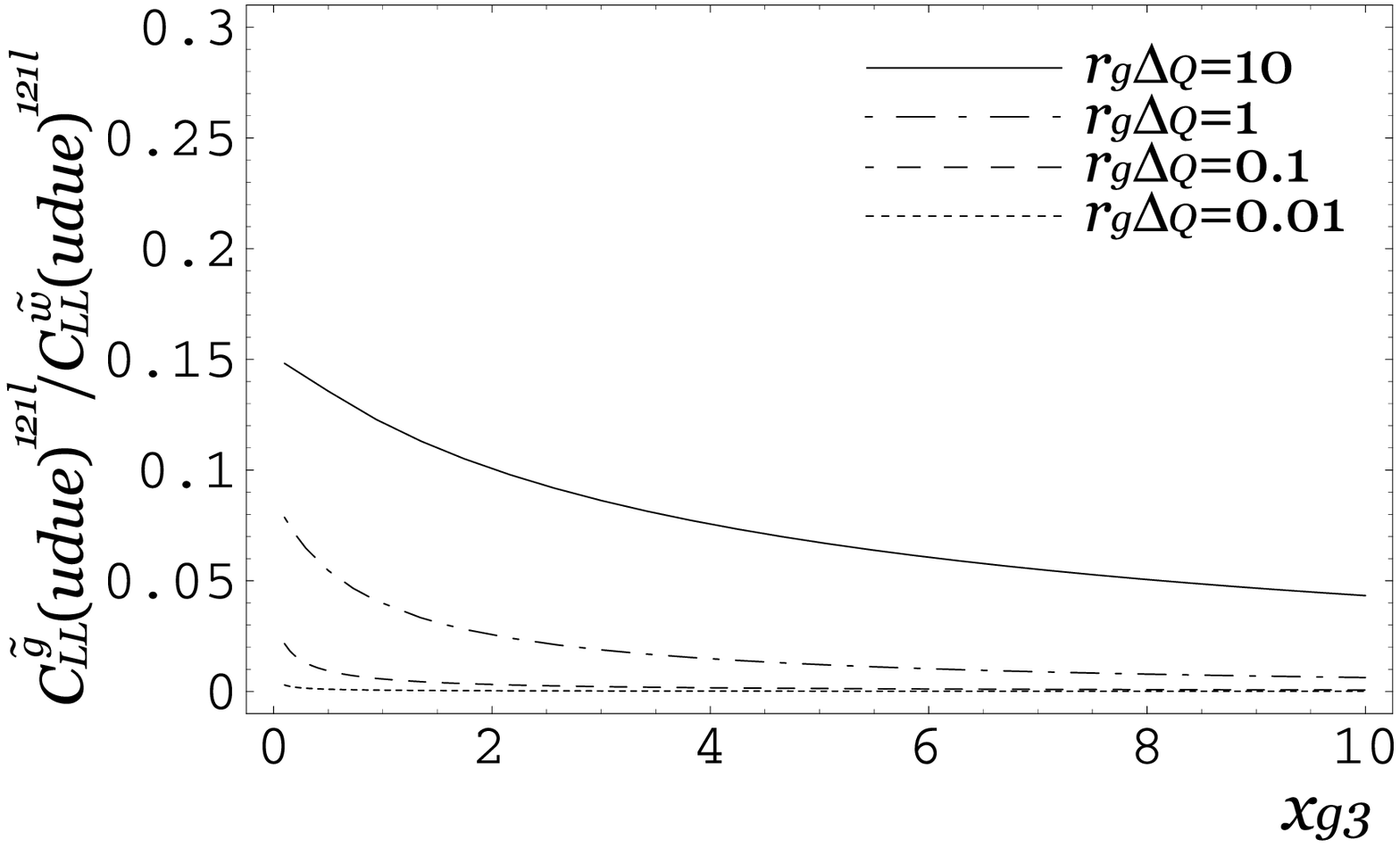} \\
  \caption{The same as  Fig.  \ref{dressingratio1}
 for the charged lepton modes.}
  \label{dressingratio2}
\end{figure}

Before we calculate the  decay amplitudes, we  investigate
the relative size of the gluino and wino  contributions.
As pointed out, 
to suppress  the SUSY contributions to FCNCs in $Q_6$ model
the degeneracy of the
first two and third generations 
needs not to be exact.
In fact, as we can see from Table III and Eq. (\ref{FCNCcondition}),
$\Delta_Q \leq 10$ is sufficient.
($\Delta_Q \equiv \Delta a_L^q$ defined in (\ref{deltaAt1}) expresses 
the degree of the degeneracy of the squark masses.)
If this is a case, the cancellation 
of the gluino  contributions is no longer 
perfect \cite{chadha},
so that they may dominate over the wino
 ones. 
In Figs. \ref{dressingratio1}-\ref{dressingratio2},
we present the results on
the ratio  $C^{\tilde g}/C^{\tilde w}$
as a function of $x_{g3}$ for 
$r_g \Delta_Q=(10,1,0.1,0.01)$ and 
$y_{w3}=10$ (right), $1$ (left).
Figs. \ref{dressingratio1} show the  results on $C(udd\nu)$'s 
which control the size of the anti-neutrino modes,
while Figs. \ref{dressingratio2} are those on
$C(udue)$'s 
for the decay amplitudes into a charged lepton.
We can see furthermore that
(except for  $C(udue)^{121l}$ shown in 
the last two figures of Figs. \ref{dressingratio2}) the gluino  contributions can become
of the same order as the  wino  ones for $x_{g_3}\sim O(1)$
if $r_g \Delta_Q \gsim 1$.
We can also see that the smaller the value of $r_g \Delta_Q$ is,
the  smaller are the gluino  contributions, as it should be.

\begin{figure}[!t]
  \centering
  \includegraphics[width=8cm]{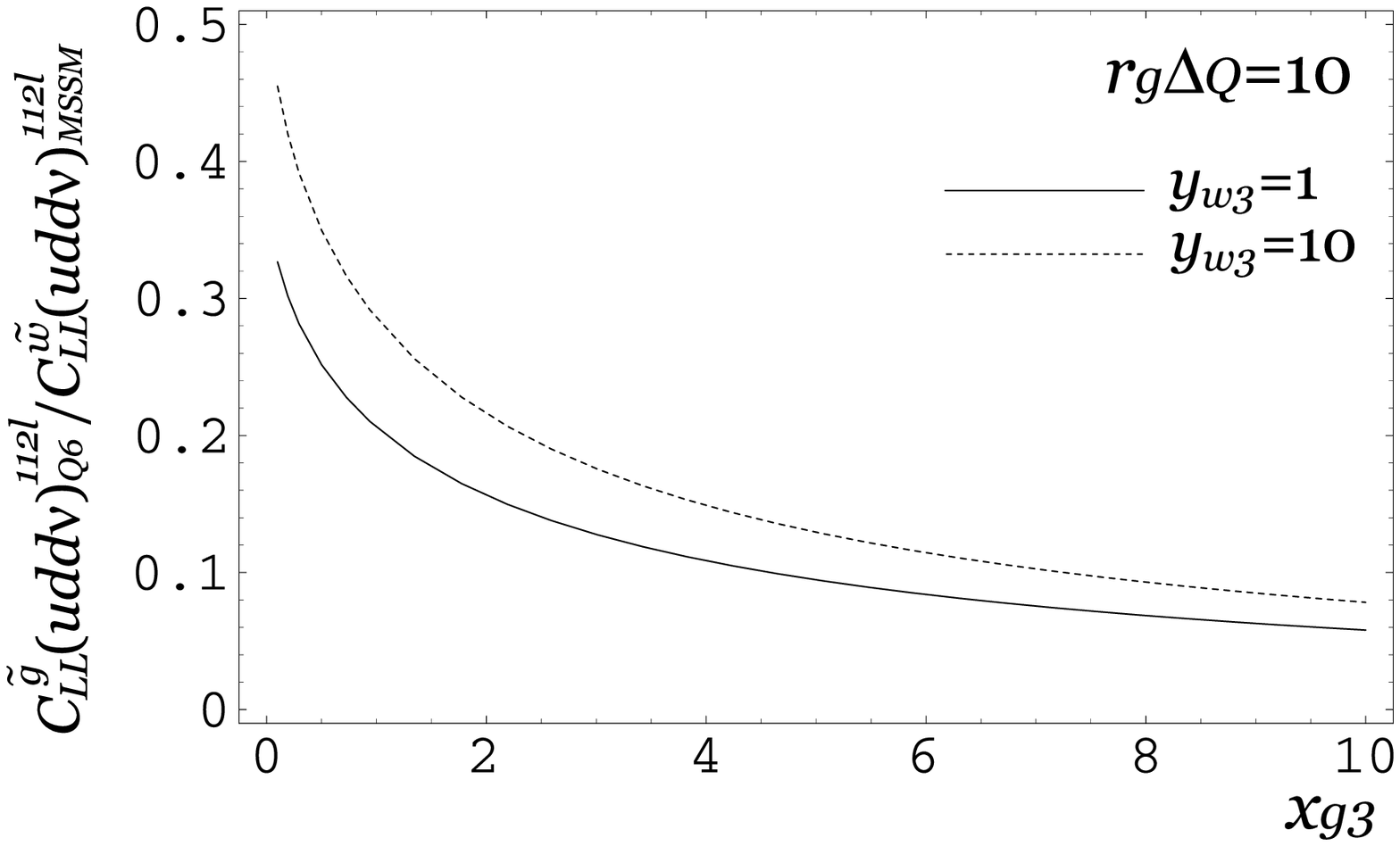}
  \caption{Ratio of $C_{LL}^{\tilde g}(udd\nu)^{112l}_{Q_6}$ to
$C_{LL}^{\tilde w}(udd\nu)^{112l}_{MSSM}$
  as a function of $x_{g3}$, where $r_g \Delta_Q=10$
  is assumed for $C_{LL}^{\tilde g}(udd\nu)^{112l}_{Q_6}$, where
$x_{g3}$ is defined in (\ref{loopfn}).
$C_{LL}(udd\nu)^{112l}$ is the largest coefficient for
  $p \to K^+ \bar \nu$, and
  the degeneracy of the squark masses ($\Delta_Q=0$) is
  assumed for $C_{LL}^{\tilde w}(udd\nu)^{112l}_{MSSM}$.}
  \label{compare}
\end{figure}

If the  degeneracy of the squark masses is assumed,
the gluino contributions vanish, and  the wino
contributions are then the dominant ones.
This is assumed in most of the previous 
investigations on  proton decay. 
Let us call the wino contribution under
the assumption of the degeneracy
(that is, $\Delta_Q=0$)   $C_{LL}(\cdots)^{\tilde w}_{MSSM}$,
and compare it with $C_{LL}(\cdots)^{\tilde g}_{Q_6}$ for which
$\Delta_Q$ may be different from zero.
Since it will turn out that
the decay mode $p \to K^+ \bar \nu$ has the largest rate, we 
restrict ourselves to $C_{LL}(udd\nu)^{112l}$
(which gives the largest contribution to the decay mode).
Fig . \ref{compare} shows the ratio
 $C_{LL}^{\tilde g}(udd\nu)^{112l}_{Q_6}/
C_{LL}^{\tilde w}(udd\nu)^{112l}_{MSSM}$
as a function of $x_{g_3}$ for $r_{g} \Delta_Q=10$
and $y_{w_3}=1,10$.
We see from the figure that the gluino contribution in 
$Q_6$ model is smaller than the wino ones of the MSSM,
although the $SU(3)_C$ gauge coupling is larger than 
that the $SU(2)_L$ gauge coupling.
This is due to the fact that  the LLLL operator in (\ref{dim5q6}) 
contains one $Q_3$ which requires a small mixing parameter
to appear when expressed in terms of the mass eigenstates of the first
two generations.

%Comparing the dominant contribution of $Q_6$ model and MSSM, we can see the effect of mixing matrix suppression.

%Then, the dominant contribution of $Q_6$ model is smaller than that of MSSM 
%even if gluino dressing contribution can not be negligible in $Q_6$ model. 
%Therefore nucleon decay rate in $Q_6$ model can be smaller than that of MSSM(FIG.(\ref{compare})).
%\begin{figure}[h]
 %\parbox{7cm}{%
%\includegraphics*[width=0.5\textwidth]{lifetimekplusnubarg-w.eps}
%\caption{partial lifetime of $p \to K^+ \bar \nu$}}
%\hspace{2cm}
 %\parbox{7cm}{%
%\includegraphics*[width=0.5\textwidth]{lifetimepiplusnubarg-w.eps}
%\caption{partial lifetime of $p\to\pi^+ \bar \nu$}}
%\end{figure}
%\begin{figure}[h]
 %\parbox{7cm}{%
%\includegraphics*[width=0.5\textwidth]{lifetimepizeroeplusg-w.eps}
%\caption{partial lifetime of $p \to \pi^0 e^+$}}
%\hspace{2cm}
% \parbox{7cm}{%
%\includegraphics*[width=0.5\textwidth]{lifetimekzeroeplusg-w.eps}
%\caption{partial lifetime of $p\to K^0e^+ $}}
%\end{figure}

\subsection{Partial decay widths and their relative size}
Here we calculate the partial decay widths of the proton.
The partial decay widths for the decay mode 
$p \to M \ell$,  $\Gamma(p \to M \ell)$,  can be written as  \cite{goto}
\be
\Gamma \left( p \to M \ell  \right)=\left(\frac{{\cal A}\beta_p}{(4 \pi)^2 
M_{PL}}\right)^2 
\frac{\left( m_P^2-m_M^2 \right)^2}{32 \pi m_P^3 f_{\pi}^2}
\left| Amp\left(  p \to M \ell \right)\right|^2,
\ee
with
\be
Amp\left(p \to K^+\bar \nu \right)&=&\kappa_6 C_{LL}(udd\nu)^{121l}+
\kappa_2 C_{LL}(udd\nu)^{112l}\\
Amp\left( p \to \pi^+ \bar \nu \right)&=&\sqrt{2}\kappa_5C_{LL}
(udd\nu)^{111l},\\
Amp\left( p \to K^0 e^+_l \right)&=&\kappa_1 C_{LL}(udue)^{121l},\\
Amp\left( p \to \pi^0 e^+_l \right)&=&\kappa_5 C_{LL}(udue)^{111l},
\ee
where $\kappa$'s  are  defined by \cite{goto}
\be
\kappa_1=1+\frac{m_P}{m_B}(F-D)=0.70&,&
~\kappa_2=1+\frac{m_P}{3 m_B}(3F+D)=1.6, \nn \\
\kappa_3=1+\frac{m_P}{m_B}(F+D)=2.0&,&~\kappa_4=2+\frac{2m_P}{m_B}F=2.7, \\ \nn
\kappa_5=\frac{1}{\sqrt{2}}(1+F+D)=1.6&,&~\kappa_6=\frac{2m_P}{3m_B}D=0.4,
\ee
and 
\be
D&=&0.81,~F=0.44,~\beta_p=0.003~{\mbox {GeV}}^3,~ 
f_{\pi}=131~{\mbox {MeV}},\nn \\
m_P&=&938~{\mbox {MeV}},~m_B=1150~{\mbox {MeV}},~m_K=495~{\mbox {MeV}},~
m_{\pi}=140~{\mbox {MeV}}.
\ee 
Here,  $D,~F$ stand for the coupling constants 
for the interaction between baryons and mesons, 
$\beta_p$ for the hadronic matrix element,
$f_{\pi}$ for the pion decay constant,
$m_P$ for  the proton mass,  $m_B$ for
the averaged baryon mass,  and $m_K$ and $m_{\pi}$
for the Kaon and  pion mass, respectively.
Further, ${\cal {A}}$ is the renormalization group(RG) enhancement factor 
for the coefficient $C_L$ \cite{rudaz,yanagida}, which 
 in our case becomes
\be
{\cal A}=10.5 ~~{\mbox{ (for the MSSM)}},~12~~{\mbox{ (for~}} 
Q_6{\mbox {~model)}}.
\ee
(Since $Q_6$ model contains four more Higgs 
doublets than in the case of the MSSM, 
the enhancement factor is slightly  larger.)

\begin{figure}[!t]
  \centering
  \hspace*{-5mm}
  \includegraphics[width=8cm]{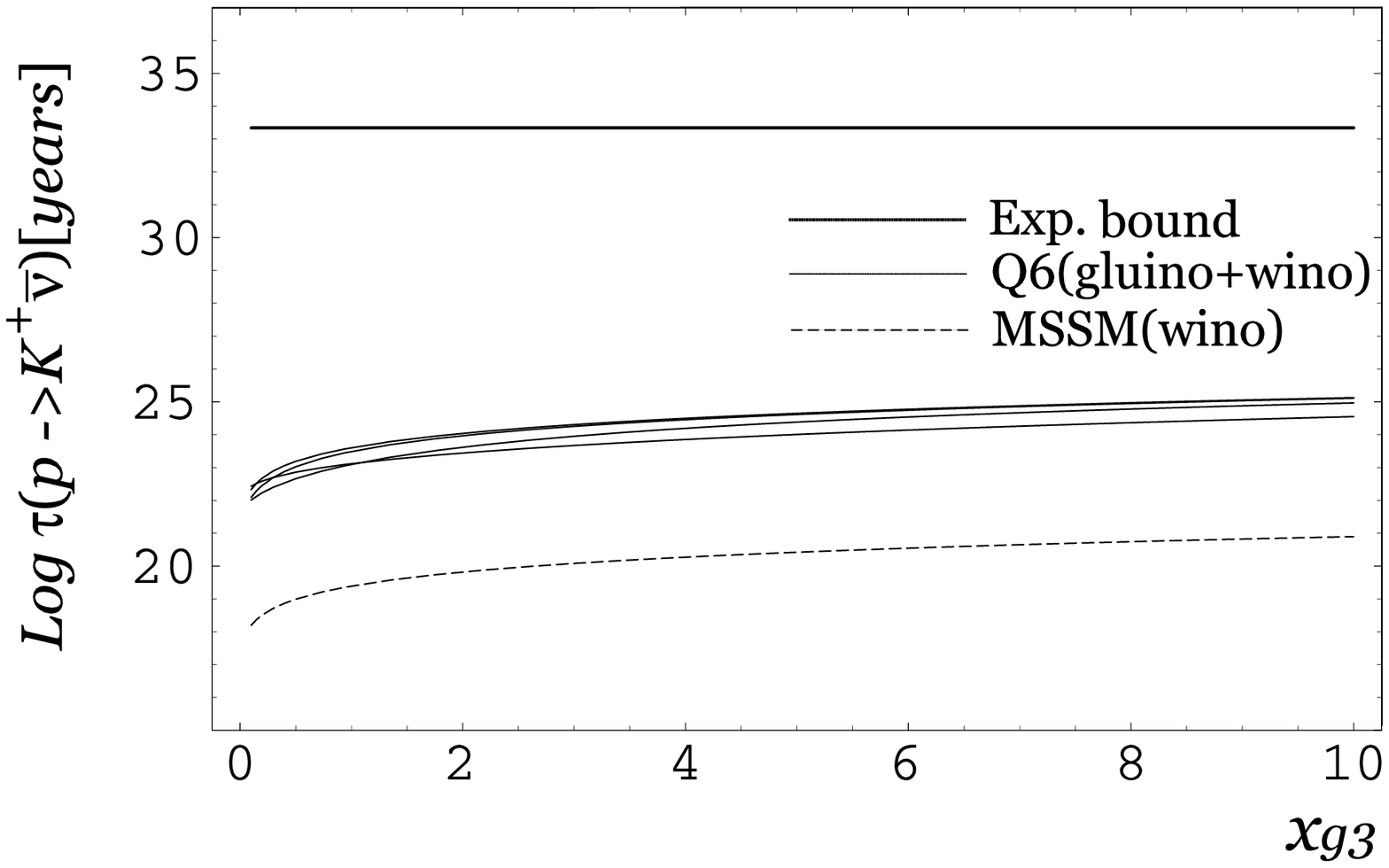}
  \hspace{5mm}
  \includegraphics[width=8cm]{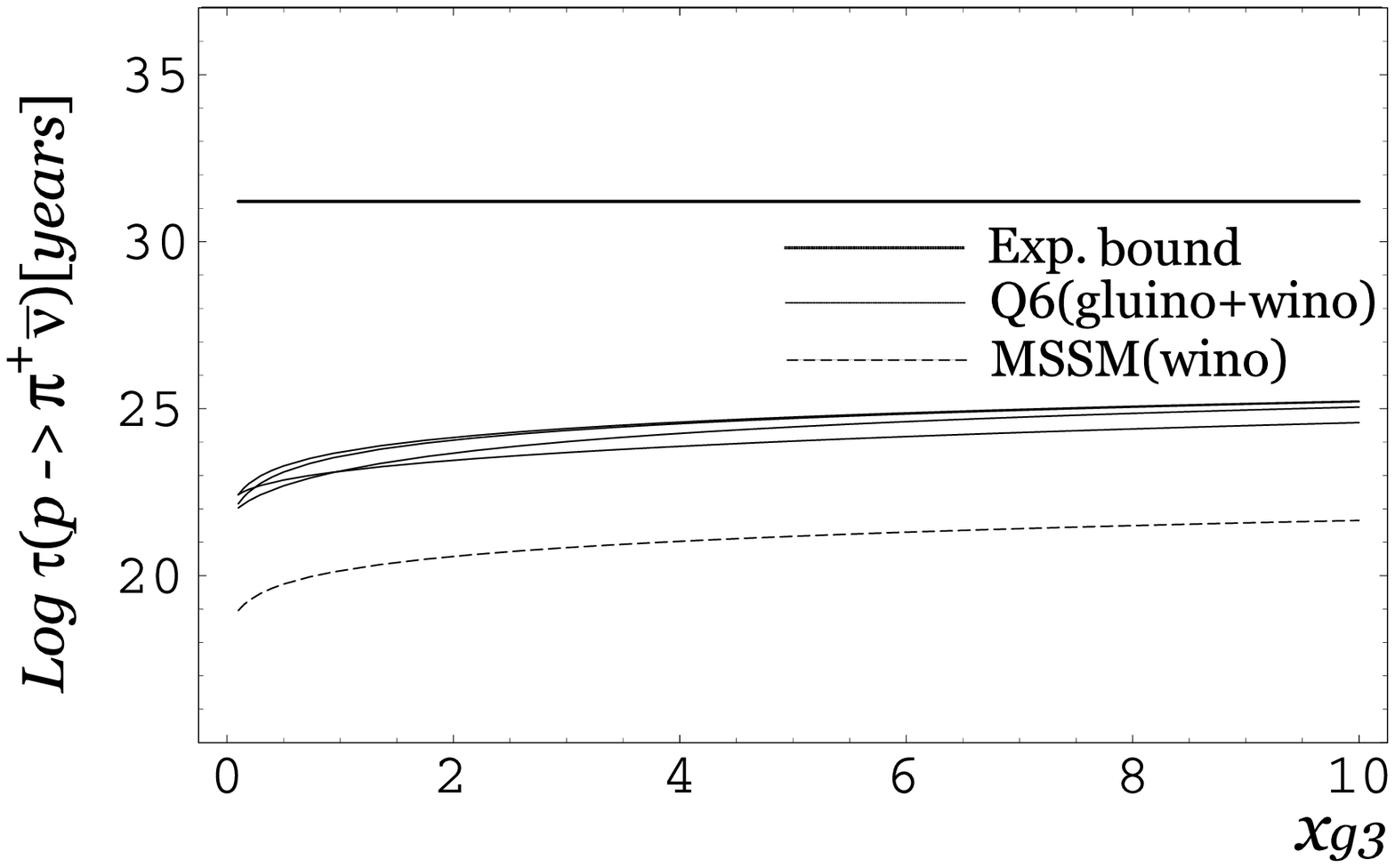} \\
\vspace{1cm}
  \hspace*{-5mm}
  \includegraphics[width=8cm]{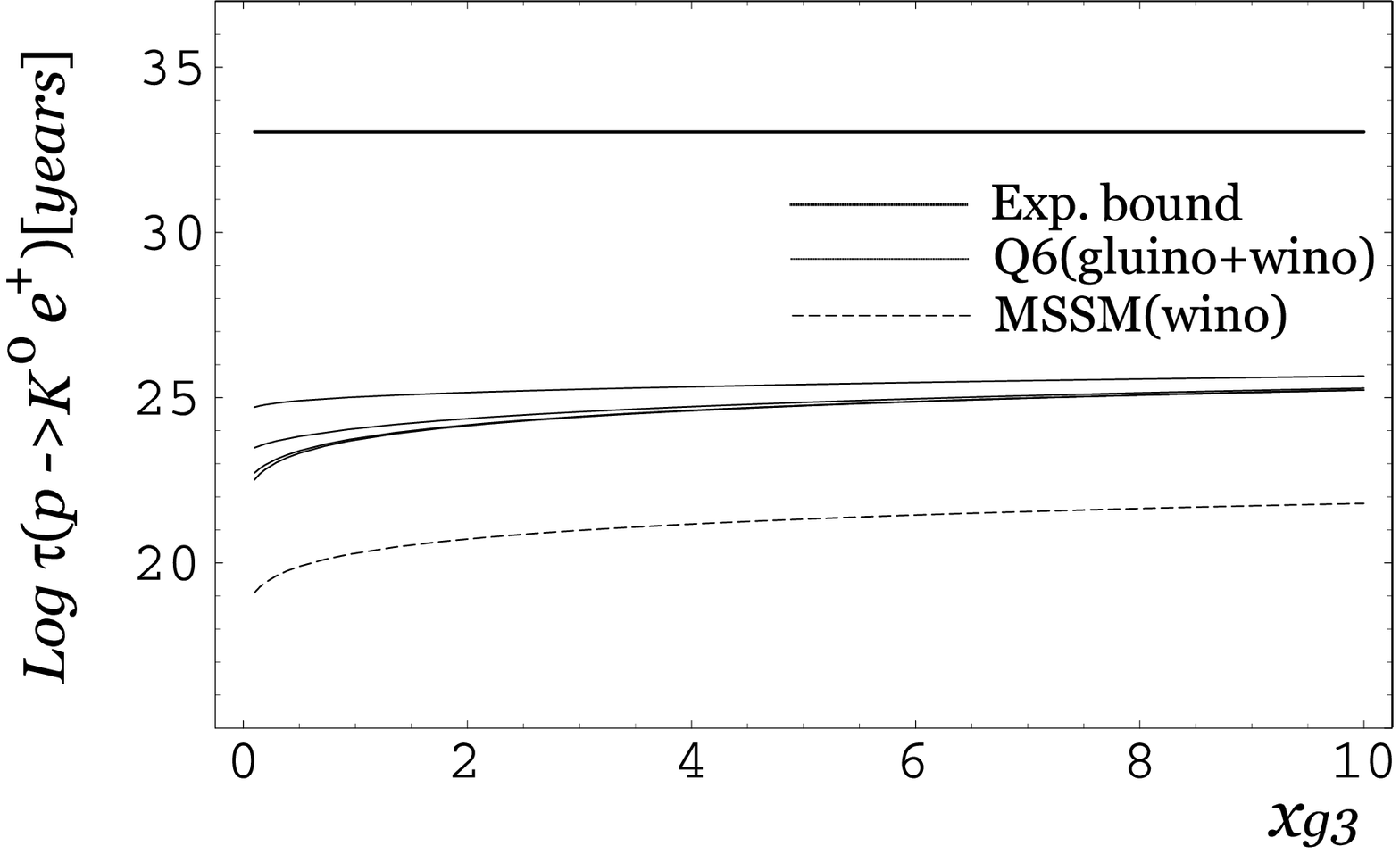}
  \hspace{5mm}
  \includegraphics[width=8cm]{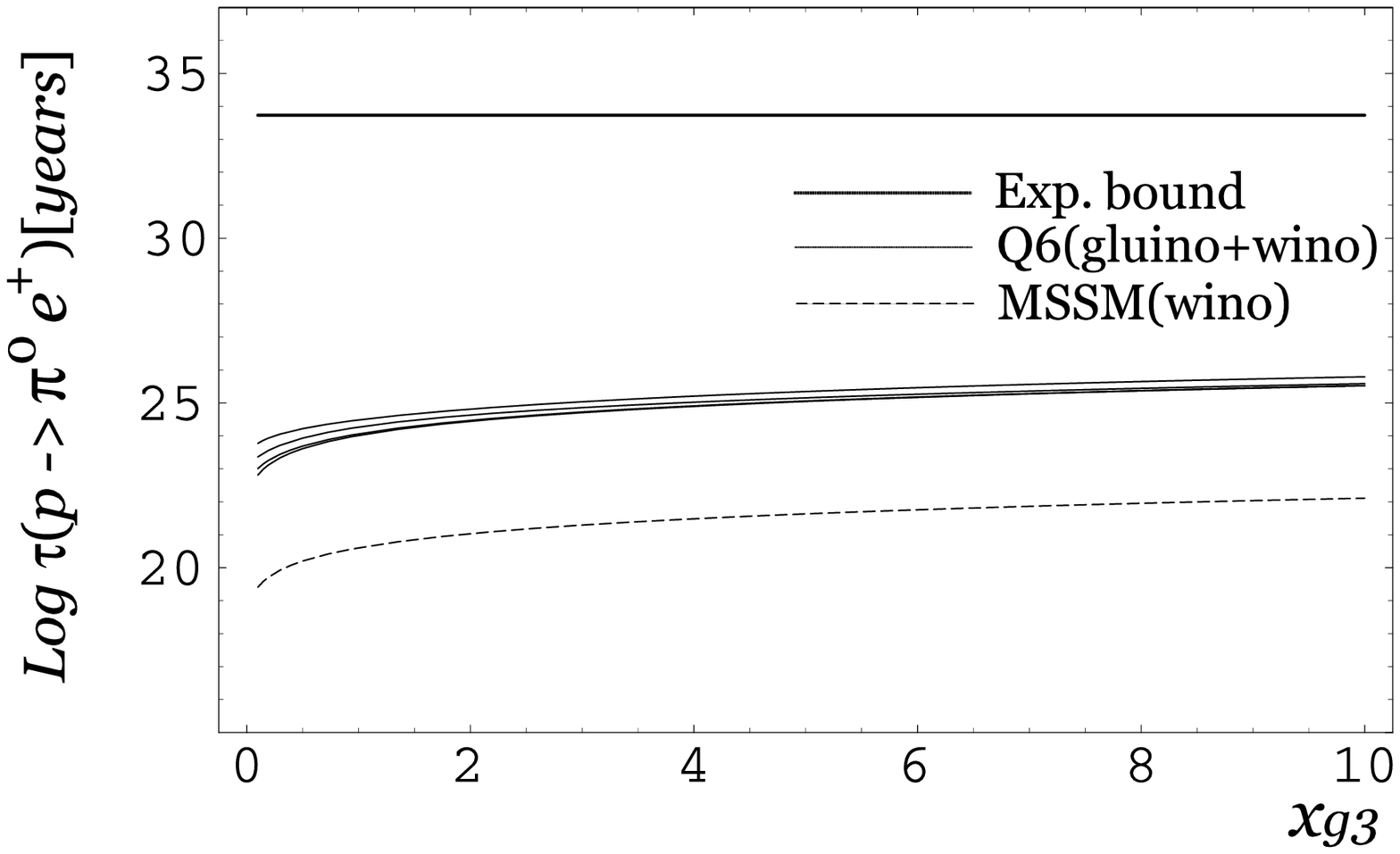} \\
  \caption{Partial lifetime of the proton for each decay mode as a function of $x_{g3}$ with $C_L=C_L^{MSSM}=1,~
  m_{\tilde g}=1~\mbox{TeV}$. 
The experimental bound (thick solid line), 
the partial lifetime calculated  in $Q_6$ model (solid line) and 
that in the MSSM (dotted line) are plotted.
The lifetime in $Q_6$ model is calculated from the sum of 
the gluino and wino contributions for $r_g\Delta_Q=0.01,0.1,1,10$, which 
corresponds to four solid lines.}
  \label{lifetime}
\end{figure}

As the first task, we would like to  compare the decay rates 
in the MSSM and $Q_6$ model.
By the MSSM decay rates we mean the decay rates which 
are obtained from the LLLL operators under the assumption
of the degenerate squark masses (no gluino dressing).
We also assume that  all the coefficients of the operators
are equal to the single constant $C_L^{MSSM}$, and that all the fields 
appearing in the operators are in the mass eigenstates
(no mixing matrix).
In  Figs. \ref{lifetime}, we plot  for each decay mode, the
experimental bound (thick solid line), 
the partial lifetime calculated  in $Q_6$ model (solid line) and 
that in the MSSM (dotted line), where we assume
\be
C_L=C_L^{MSSM}=1,~~m_{\tilde q}=1~{\mbox{TeV}}.
\ee   
The lifetime in $Q_6$ model is calculated from the sum of 
the gluino and wino contributions for $r_g\Delta_Q=0.01,0.1,1,10$, which 
corresponds to
four solid lines. We first  see from the figures that
if $r_g\Delta_Q$ varies from $0.01$ to $10$,
a change of the life time about an order of magnitude 
can appear.
We also see that the decay rates in $Q_6$ model is much more suppressed
than those in the MSSM.
Quantitatively, we
 find that the experimental bounds can be satisfied if
 the coefficient for the LLLL operator satisfies
\be
C_L < 10^{-(4 \sim 5)},
\ee 
while  $C_L^{MSSM}<10^{-(6 \sim 7)}$
should be satisfied 
for the case of the MSSM. That is, $Q_6$ flavor symmetry
can suppress  proton decay by
four orders of magnitude (which can be  also seen from the figures).

\begin{figure}[!t]
  \centering
  \hspace*{-5mm}
  \includegraphics[width=8cm]{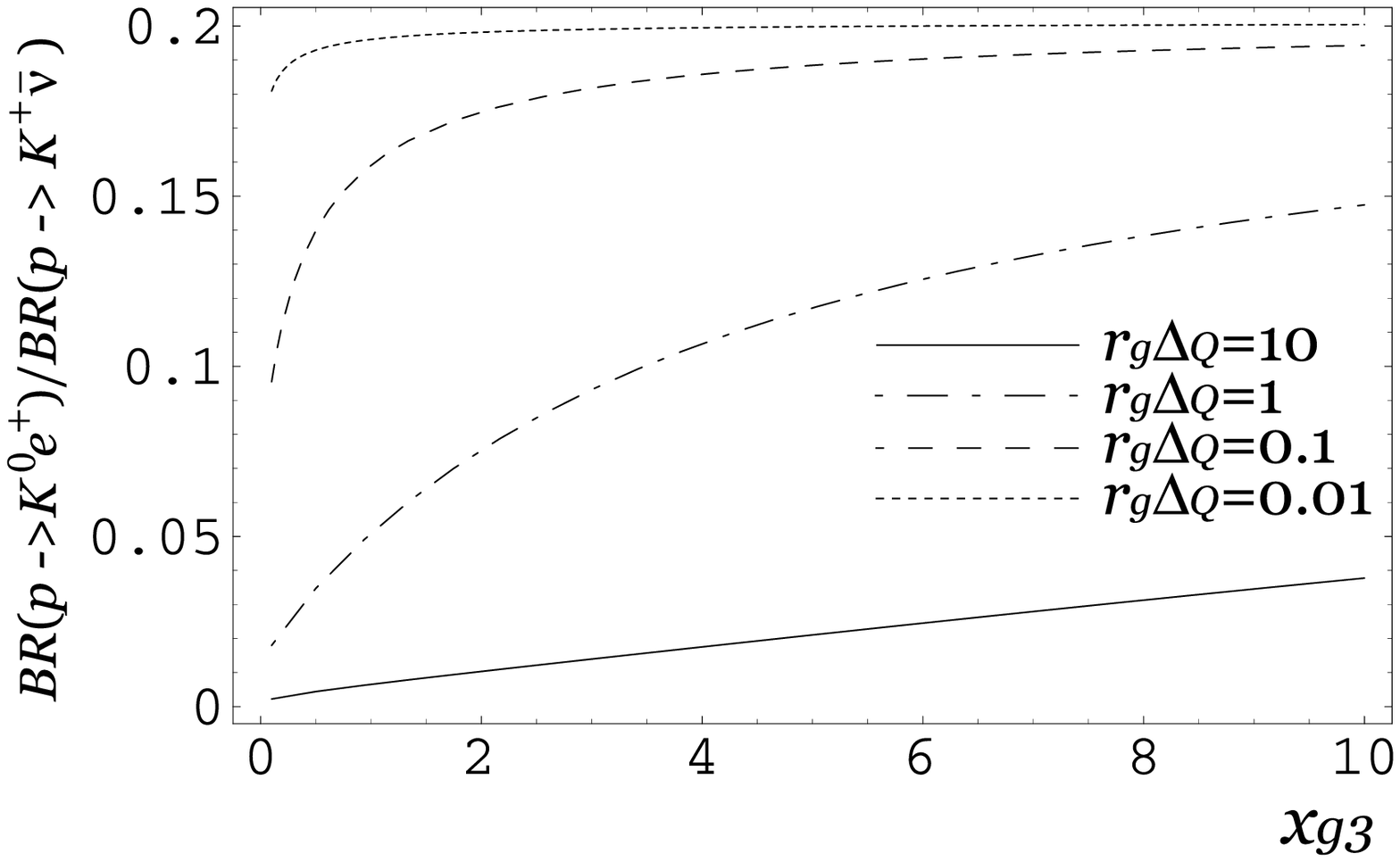}
  \hspace{5mm}
  \includegraphics[width=8cm]{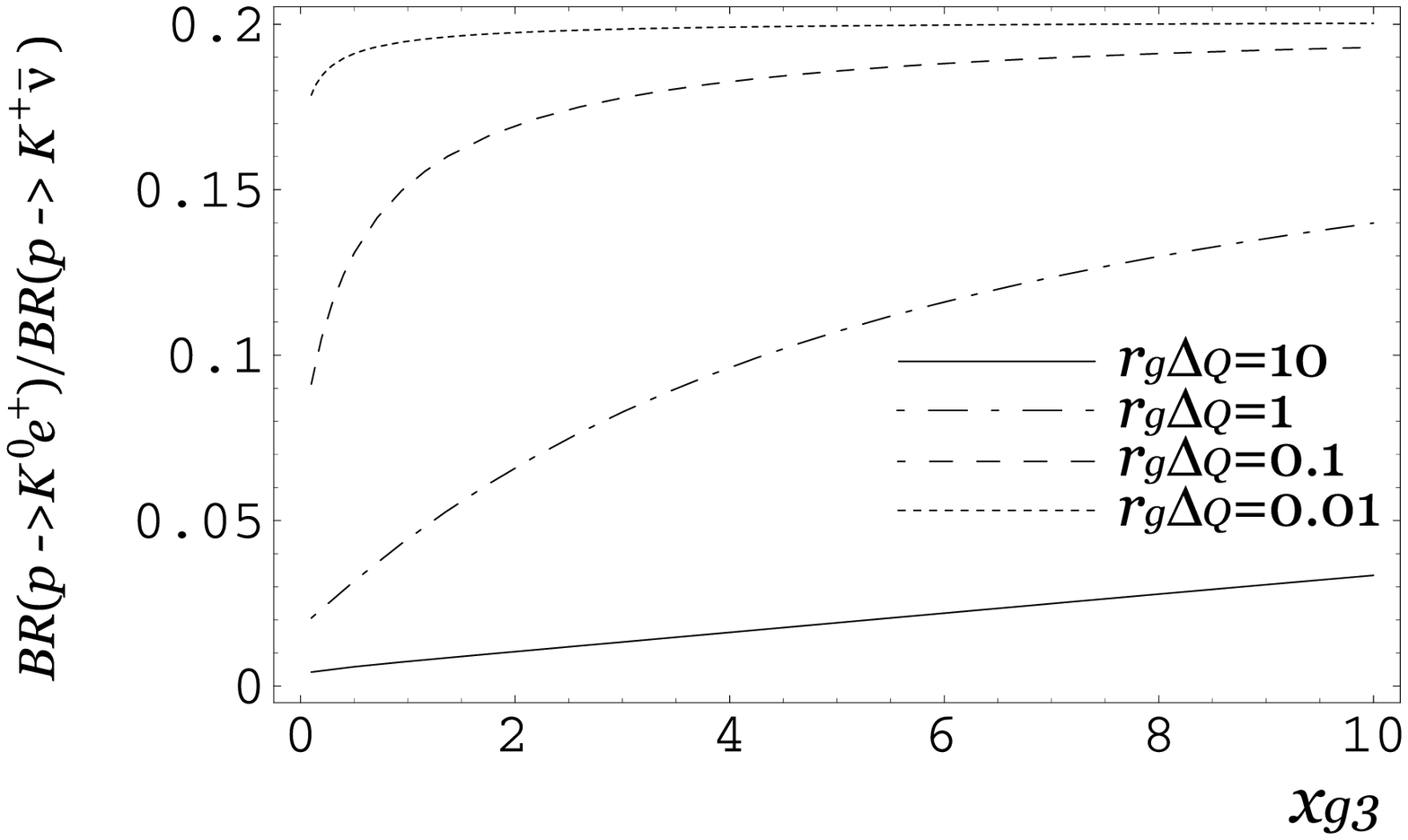} \\
\vspace{1cm}
  \hspace*{-5mm}
  \includegraphics[width=8cm]{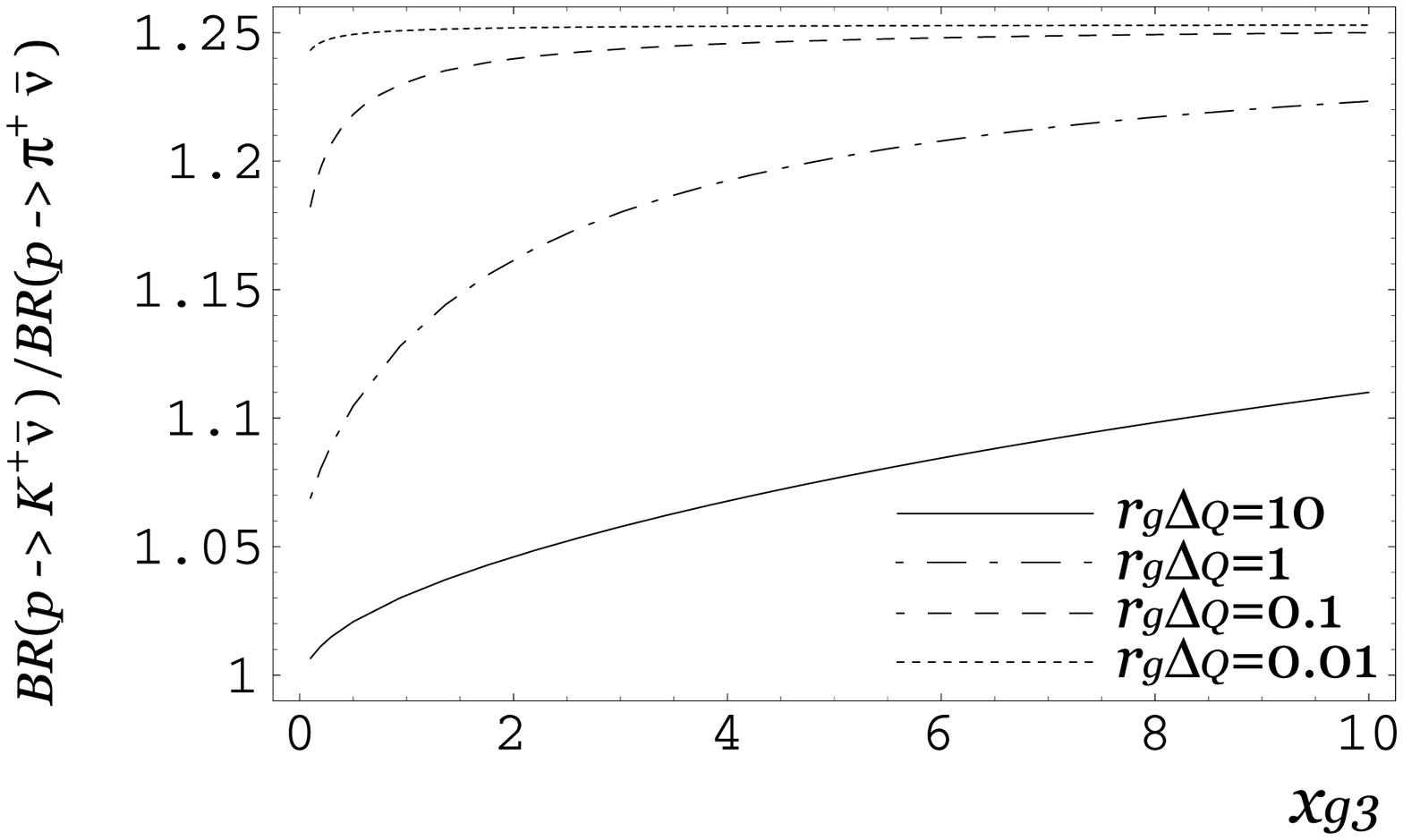}
  \hspace{5mm}
  \includegraphics[width=8cm]{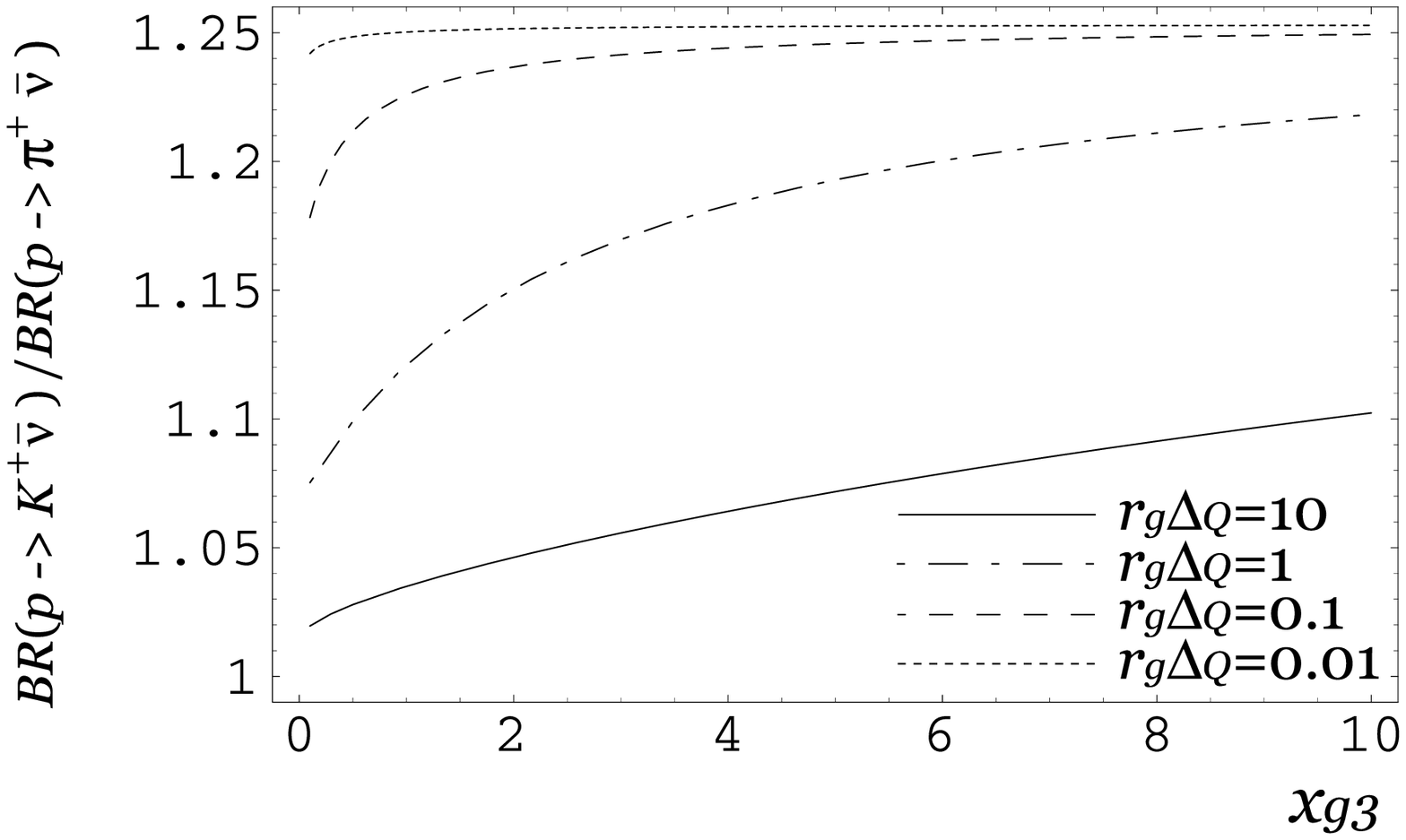} \\
  \caption{Ratio of partial decay rates for each decay mode with $r_g\Delta_Q=(10,~1,~0.1,~0.01)$.
  Figures in the left column are for $y_{w3}=1$, and the right column for $y_{w3}=10$.}
  \label{ratio1}
\end{figure}

\begin{figure}[!t]
  \centering
  \hspace*{-5mm}
  \includegraphics[width=8cm]{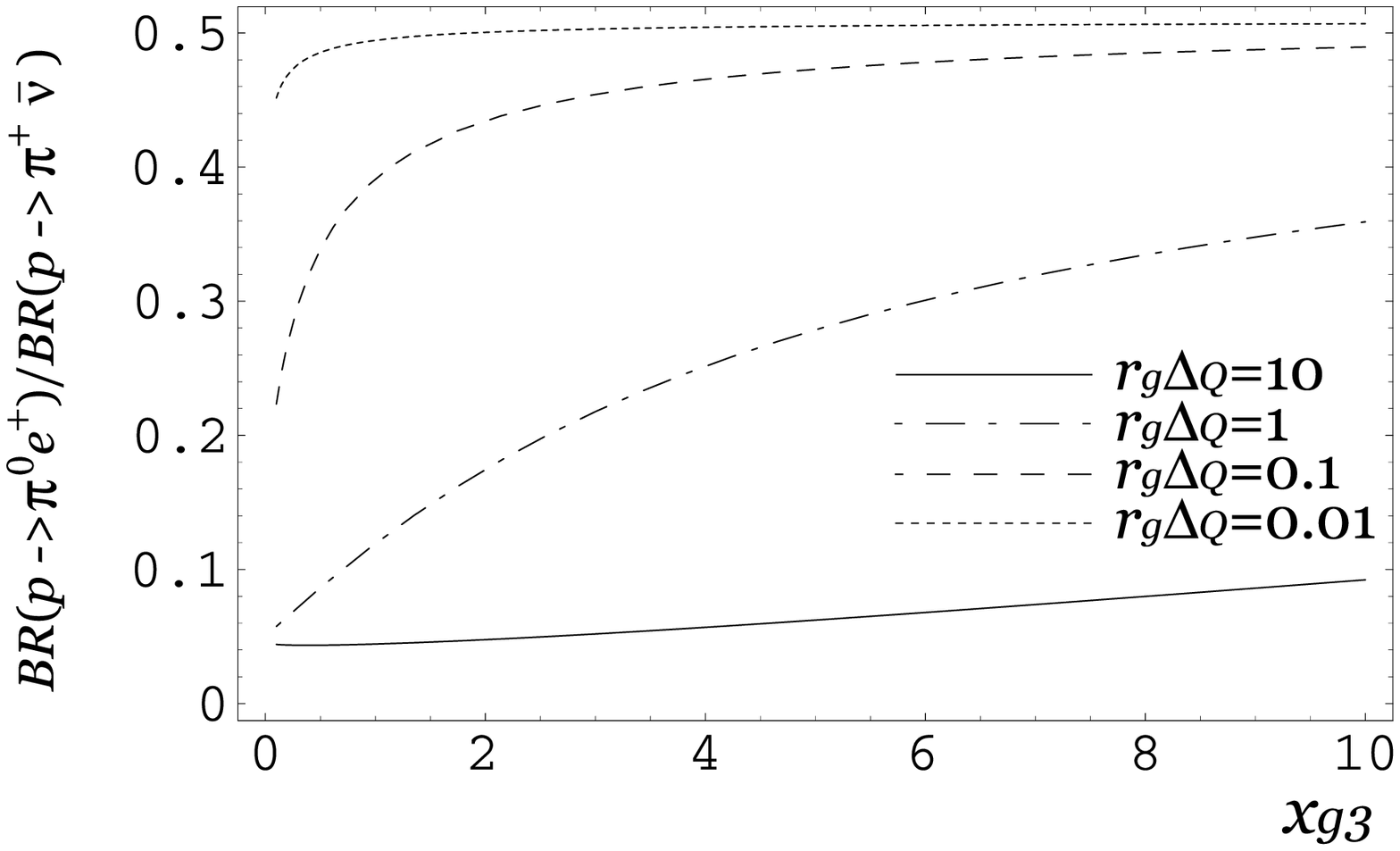}
  \hspace{5mm}
  \includegraphics[width=8cm]{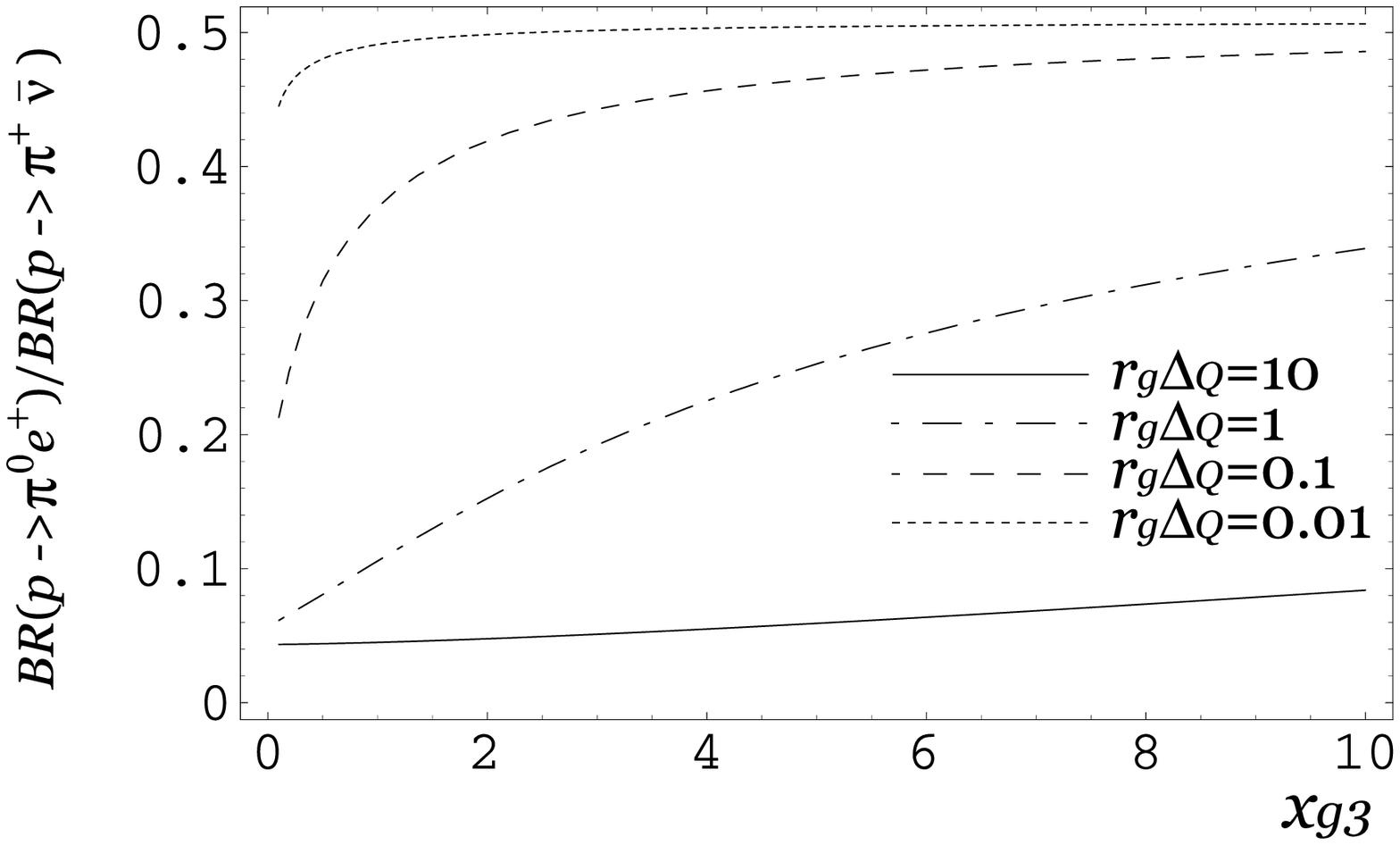} \\
\vspace{1cm}
  \hspace*{-5mm}
  \includegraphics[width=8cm]{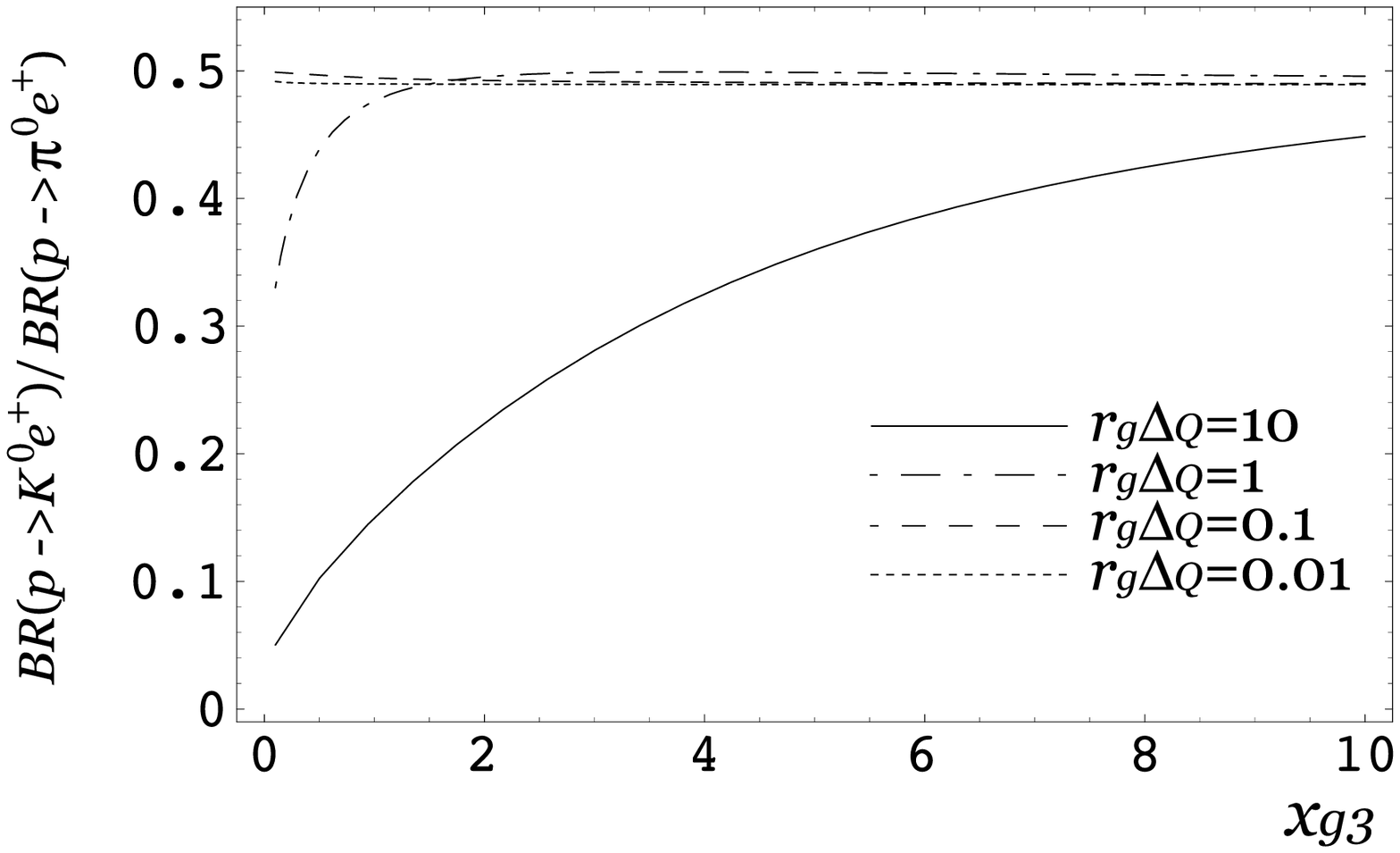}
  \hspace{5mm}
  \includegraphics[width=8cm]{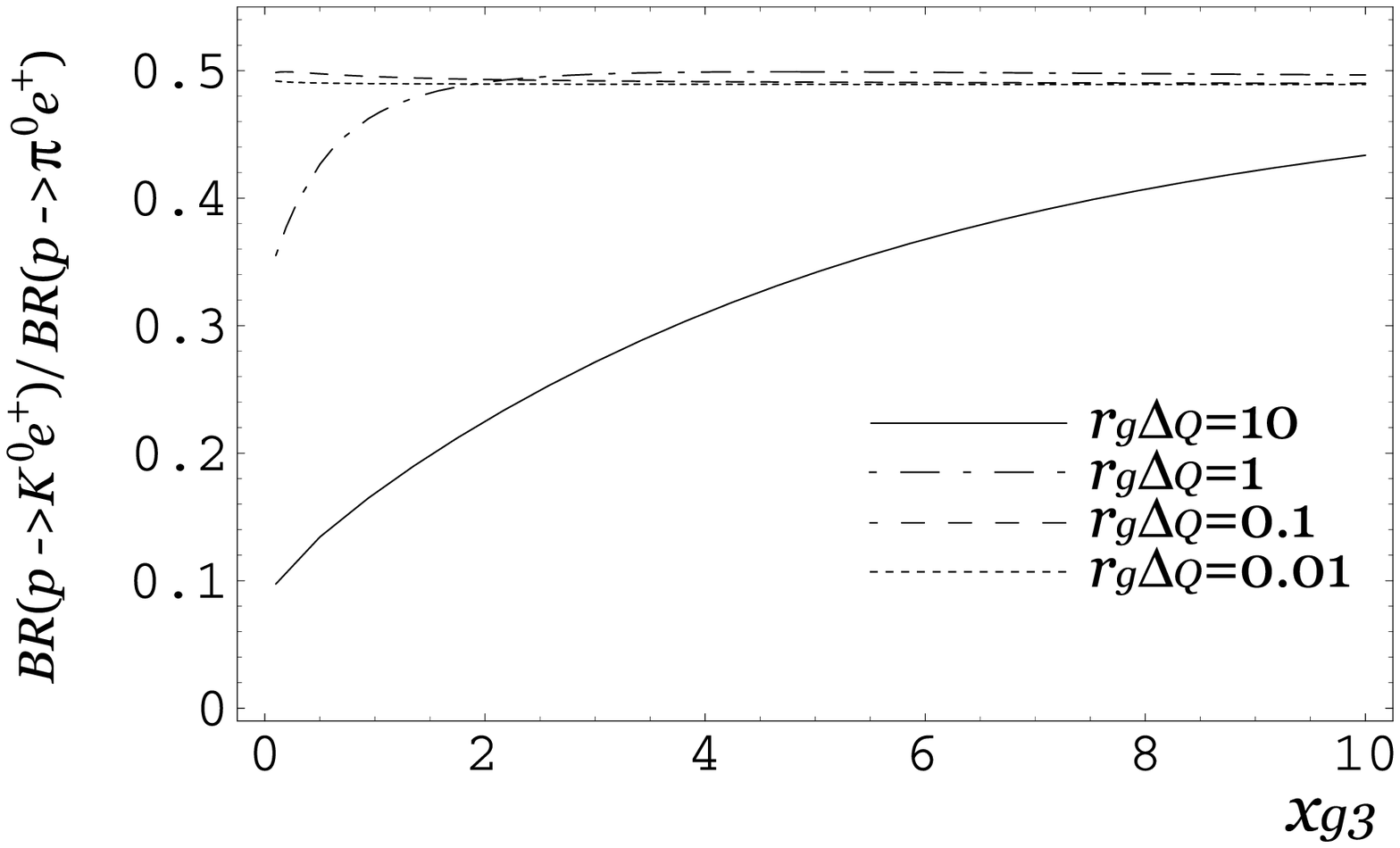} \\
  \caption{Ratio of partial decay rates for each decay mode with $r_g\Delta_Q=(10,~1,~0.1,~0.01)$.
  Figures in the left column are for $y_{w3}=1$, and the right column for $y_{w3}=10$.}
  \label{ratio2}
\end{figure}

Next we would like to compare
our results with those obtained in  the minimal SUSY $SU(5)$ GUT.
In our lowest order approximation, 
only $C_L$ is an independent
coefficient, implying that
the ratio of  partial decay widths is independent of
$C_L$. That is, in $Q_6$ model, the relative 
size of the partial decay rates is fixed (once the SSB sector is fixed).
First we recall the case of the minimal SUSY $SU(5)$ GUT 
\cite{yanagida,murayama}.
The superpotential for the baryon
number violating effective dimension-five operators in this case is given by
\be
W_5^{SU(5)}=\frac{1}{2M_{H_C}}\left[ y_u y_{dl} \left( V^*\right)_{2l}
\left(Q_1 Q_1 \right)
\left( Q_2 L_l \right)+y_c y_{dl} \left( V^*\right)_{1l}\left(Q_2 Q_2 \right)
\left( Q_1 L_l \right)\right]
\label{w5su5}
\ee
in the approximation that the third generation is dropped,
where $y_{u,c}, y_{dl}$ are the diagonal Yukawa 
couplings of the corresponding quarks, $l=1,2$ is generation index
(i.e., $d1=d$ and $d2=s$), and  $V$ is the CKM matrix. 
In writing (\ref{w5su5}), a nontrivial assumption is made;
the up quark Yukawa matrix is diagonal over the whole range of
energies. Therefore, the superpotential (\ref{w5su5})
is not a unique prediction of the the model.
Under the assumption of the degeneracy of the squark masses, 
 only the wino dressing diagrams contribute to the decay,
 and one finds \cite{yanagida,murayama}
\be
\Gamma \left( p \to K^+ \bar \nu \right)&=&\gamma 
\left( m_P^2-m_K^2 \right)^2 
\left| y_c y_s \sin^2 \theta_C \kappa_3 \right|^2, \\ 
\Gamma \left( p \to \pi^+ \bar \nu \right)&=&\gamma 
\left( m_P^2-m_{\pi}^2 \right)^2 
\left| 2 y_c y_s \sin^2 \theta_C \tan \theta_C \sqrt{2}
\kappa_5 \right|^2, \\
\Gamma \left( p \to K^0 e^+_l \right)&=&\gamma 
\left( m_P^2-m_K^2 \right)^2 
\left| y_u y_{dl} \left( V^* \right)_{2l} \cos \theta_C 
\kappa_1 \right|^2, \\
\Gamma \left( p \to \pi^0 e^+_l  \right)&=&
\gamma \left( m_P^2-m_{\pi}^2 \right)^2 
\left| y_u y_{dl} \left( V^*\right)_{2l} \sin \theta_C 
\kappa_5 \right|^2, 
\ee
where we have dropped the corresponding loop functions (\ref{loopfn}), and
 the common factor $\gamma$ is given by
\be
\gamma=\left( \frac{{\cal A}\beta_p \alpha_2 
\cos \theta_C}{\pi M_{H_C}}\right)^2 
\frac{1}{32 \pi m_P^3 f_{\pi}^2}.
\ee
 From these partial decay widths, we obtain
the  relative decay widths in the minimal $SU(5)$ 
GUT \cite{yanagida,murayama}: 
\be
\frac{B\left( p \to K^+ \bar \nu \right)}{B\left(p \to \pi^+ 
\bar \nu  \right)}&=& 
\frac{ \left( m_P^2-m_K^2 \right)^2 }{\left( m_P^2-m_{\pi}^2 \right)^2 }
\left( \frac{\kappa_3}{2 \sqrt{2}\kappa_5 \tan \theta_C }
\right)^2 \simeq 2, \\
\frac{B\left( p \to K^0 e^+_l \right)}{B\left(p \to \pi^0 
e^+_l  \right)}&=&
\frac{ \left( m_P^2-m_K^2 \right)^2 }{\left( m_P^2-m_{\pi}^2 \right)^2 }
\left( \frac{\kappa_1 \cos \theta_C}{\kappa_5 \sin \theta_C}
 \right)^2 \simeq 2, \\
\frac{B\left( p \to K^0 \mu^+ \right)}{B\left(p \to K^+ 
\bar \nu  \right)}&=&
\left( \frac{y_u \kappa_1 \cos \theta_C }{y_c \kappa_3 
\sin^2 \theta_C }\right)^2
\simeq 6 \times 10^{-4}, \\
\frac{B\left( p \to \pi^0 \mu^+ \right)}{B\left(p \to \pi^+ 
\bar \nu  \right)}&=&
\left( \frac{y_u \cos^2 \theta_C }{2 \sqrt{2}y_c
  \sin^2 \theta_C }\right)^2
\simeq 3 \times 10^{-4}.
\ee

The corresponding results for $Q_6$ model are found to be 
\be
\frac{B\left( p \to K^+ \bar \nu \right)}{B\left(p \to \pi^+
 \bar \nu  \right)}&=& 
\frac{ \left( m_P^2-m_K^2 \right)^2 }{\left( m_P^2-
m_{\pi}^2 \right)^2 }
\left( \frac{ \kappa_2 U_{dL}^{32}}{\sqrt{2}\kappa_5
 U_{dL}^{31}} \right)^2
\simeq 
\left\{\begin{array}{l}
1.25 \\
1\\
\end{array}\right. , \label{nunu} \\
\frac{B\left( p \to K^0 e^+_l \right)}{B\left(p \to \pi^0 
e^+_l  \right)}&=&
\frac{ \left( m_P^2-m_K^2 \right)^2 }{\left( m_P^2-
m_{\pi}^2 \right)^2 }
\left\{ \begin{array}{l}
\left( \frac{\kappa_1 U_{dL}^{32}}{\kappa_5 U_{dL}^{31}}
\right)^2 \\
\left( \frac{\kappa_1 U_{uL}^{31} \sin \theta_C}{\kappa_5 
U_{uL}^{31}}\right)^2\\
\end{array}\right. \simeq
\left\{ \begin{array}{l} 
0.5 \\
5 \times 10^{-3}\\
\end{array}\right.,  \label{ellell} \\
\frac{B\left( p \to K^0 e^+_l \right)}{B\left(p \to K^+ 
\bar \nu  \right)}&=&
\left( \frac{\kappa_1}{\kappa_2}\right)^2 \left| U_{eL}^{3 l} \right|^2
\left\{ \begin{array}{l}
1 \\
\left( \frac{U_{uL}^{31}\sin \theta_C}{U_{dL}^{32}}\right)^2 \\
\end{array}\right. \simeq
\left| U_{eL}^{3l} \right|^2
\left\{ \begin{array}{l} 
0.2 \\
2.4 \times 10^{-4} \\
\end{array} \right., \label{kellnu} \\
\frac{B\left( p \to \pi^0 e^+_l \right)}{B\left(p \to \pi^+
 \bar \nu  \right)}&=&
\frac 12 \left| U_{eL}^{3 l} \right|^2
\left\{ \begin{array}{l}
1\\
\left( \frac{U_{uL}^{31}}{U_{dL}^{31}}\right)^2 \\
\end{array}\right.  \simeq
\left| U_{eL}^{3 l} \right|^2
\left\{ \begin{array}{l} 
0.5 \\
5.1 \times 10^{-2} \\
\end{array} \right., \label{piellnu}
\ee
where as in the case of the minimal SUSY GUT we have suppressed 
the loop functions. The upper (lower)  numbers on the right hand side 
correspond to the wino (gluino) contributions.
Note that in contrast to the case of the minimal SUSY GUT
the mixing parameters  explicitly appear,  reflecting
the flavor structure of the present $Q_6$ model.
The most remarkable difference is the charged lepton modes.
As we see from (\ref{kellnu}) and (\ref{piellnu}), they are proportional to
$|U_{eL}^{3l}|^2$, where the unitary matrix $U_{eL}$
(which rotates the left-handed charged leptons) is explicitly 
given in (\ref{UeL}). We find
\be
\frac{B\left( p \to K^0 \mu^+ \right)}{B\left( p \to K^0 e^+ \right)}
&=&\frac{B\left( p \to \pi^0 \mu^+ \right)}{B\left( p \to \pi^0 e^+ \right)}
=\frac{\left| U_{eL}^{32}\right|^2}{\left| U_{eL}^{31}\right|^2}\nn\\
&=& \left(\frac{m_e}{m_\mu}\right)^2
\simeq 2.37 \times 10^{-5}. 
\ee
The same ratio in the case of the minimal SUSY GUT becomes 
$B\left( p \to K^0 \mu^+ \right)\simeq 10^3 
\times B\left( p \to K^0 e^+ \right)$
if we use the superpotential (\ref{w5su5}).
In Figs. \ref{ratio1} and \ref{ratio2}
we plot four different combinations of ratios
as a function of $x_{g3}$ for various values of
$r_g \Delta_Q$  in $Q_6$ model.
The figures on the left (right) side correspond to $y_{w_3}=1 (10)$. From 
the figures, we  find the hierarchical structure of the decay modes:
\be
B\left(p \to K^0 e^+\right)< B\left(p \to \pi^0 e^+\right)
<B\left(p \to \pi^+ \bar \nu \right)
< B\left(p \to K^+ \bar \nu \right).
\ee
In the case of the minimal SUSY GUT we obtain instead
\be
 B\left(p \to \pi^0 e^+\right) < B\left(p \to K^0 e^+\right)
<< B\left(p \to \pi^+ \bar \nu \right)
<B\left(p \to K^+ \bar \nu \right).
\ee
Note, however, that although $B\left(p \to \pi^0 e^+\right)
<  B\left(p \to K^+ \bar \nu \right)$, they are basically in the same oder
in the $Q_6$ model. That is, the $Q_6$ model predicts that once
the decay mode $K^+ \bar \nu$ is experimentally observed,
then it is likely to observe the decay mode $\pi^0 e^+$, too,
in sharp contrast to the case of the minimal SUSY $SU(5)$ GUT.

\section{Conclusion}
Flavor symmetry can play important rolls in supersymmetric
 models \cite{bilgin}. We have investigated
the SUSY contributions 
to FCNCs and  to proton decay in a supersymmetric 
extension of the SM  based on a binary dihedral family group $Q_6$.
We have seen that 
the discrete low energy flavor symmetry  $Q_6$
can be an alternative to the universality assumption of the 
soft supersymmetry breaking parameters.
That is, the existence of a hidden sector, in which supersymmetry is 
assumed to be broken in a flavor blind manner, is not
indispensable in this model.
Therefore, a variety of supersymmetry breaking mechanisms
may become phenomenologically viable.

It has turned out from our analysis on FCNCs that
the degeneracy
requirement of the soft scalar masses of the
left-handed squarks  in this model can be
significantly relaxed. 
This has a considerable effect on the
gluino-mediated one-loop amplitudes on  proton decay.
In most of the previous calculations, the degeneracy
was assumed so that the gluino contributions are cancelled
with each other, which implies that the wino contributions are the
most dominant contributions. 
We have fond that  the non-degenerate squarks
 can change the decay rate
in the charged lepton modes by an order of magnidude.
We have also found that in the lowest order
approximation there is only one independent
coefficient for dimension-five operators
that lead to proton decay.
Consequently,
the relative partial decay rates in this approximation are fixed,
reflecting
the flavor structure dictated by $Q_6$.

Our main finding is that  $Q_6$ flavor symmetry 
acts in such a way that
 the smallness of $(V_{\rm MNS})_{e3}$, the suppression of
 $\mu \to e+\gamma$, and the smallness of
 $B\left( p \to K^0(\pi^0) +\mu^+ \right)/
 B\left( p \to K^0(\pi^0) +e^+ \right)$
 have the same origin; the electron is much lighter than 
 the muon and the tau.
 They in fact vanish in the $m_e \to 0$ limit.
 
\vspace{0.5cm}
\noindent
{\large \bf Acknowledgments}\\
We would like to thank J.~Hisano,  H.~Nakano,  T.~Singai
and K.~Tobe for useful discussions. 
E.I is supported by Research Fellowship of the Japan Society for the Promotion of Science (JSPS) for Young Scientists 
(No.16-07971).
This work is supported by the Grants-in-Aid for Scientific Research 
from the Japan Society for the Promotion of Science
(\# 13135210).

\end{document}